\documentclass[useAMS,usenatbib]{mn2e}
\usepackage{amssymb}
\usepackage{deluxetable}

\title[VLBI images and VLA polarization in CSS sources]{A sample of
  small size compact steep-spectrum radio sources. VLBI images and VLA
  polarization at 5 GHz}
\author[D. Dallacasa et al. ]
  {D. Dallacasa$^{1,2}$\thanks{E-mail: ddallaca@ira.inaf.it},
M. Orienti$^{1,2}$, C. Fanti$^{2}$, R. Fanti$^{2}$, C. Stanghellini$^{2}$\\
$^1$Dipartimento di Astronomia, Universit\`a di Bologna, via Ranzani 1,
I-40127, Bologna, Italy \\
$^2$INAF -- Istituto di Radioastronomia, via Gobetti 101, I-40129, Bologna,
Italy \\
}
\date{Received \today; accepted ?}

\pagerange{\pageref{firstpage}--\pageref{lastpage}} \pubyear{2002}

\def\LaTeX{L\kern-.36em\raise.3ex\hbox{a}\kern-.15em
    T\kern-.1667em\lower.7ex\hbox{E}\kern-.125emX}

\begin{document}

\label{firstpage}

\maketitle

\begin{abstract}
Global VLBI observations at 5 GHz have been performed to study
the source morphology in 10 compact steep-spectrum (CSS) sources
selected from the Peacock \& Wall catalogue with the aim of finding
asymmetric structures produced by the interaction with the ambient
medium. The combination of these data and earlier
1.7-GHz observations allows the study of the spectral
index distribution across the source structure and the unambiguous
determination of the nature of each component. 
In seven sources we detected the core component 
with a flat or inverted spectrum. In six sources the radio
emission has a two-sided morphology and comes mainly from 
steep-spectrum extended structures, like lobes, jets,
and hotspots. Only one source, 0319+121, has a
one-sided core-jet structure. In three out of the six sources with a 
two-sided structure the flux
density arising from the lobes is asymmetric, and the brightest
lobe is the one closest to the core, suggesting that the jets are
expanding in an 
inhomogeneous ambient medium which may influence the source
  growth. The interaction between the jet and the environment may slow
  down the source expansion and 
  enhance the luminosity due to severe radiative losses, likely
  producing an excess of CSS radio sources in flux density limited samples. 
The lobes of the other three asymmetric
sources have a
brighter-when-farther behaviour, in agreement with what is expected by
projection and relativistic effects. 
Simultaneous VLA observations carried out
to investigate the polarization properties of the targets detected
significant polarized emission ($\sim$5.5\%) only from the quasar 0319+121.

\end{abstract}

\begin{keywords}
galaxies: active -- galaxies: jets -- galaxies: nuclei -- 
quasars: general -- radio continuum: galaxies -- radio
continuum: general
\end{keywords}

\section{Introduction}

The evolutionary stage of a powerful extragalactic radio source is
currently thought to be related to its linear size. Statistical
studies of the population of radio sources in the first stages of
their individual evolution are fundamental for a comprehensive
understanding of the radio emission phenomenon and its duty-cycle.\\
Compact steep-spectrum (CSS) and Gigahertz-peaked spectrum (GPS)
radio sources  
are intrinsically small-sized (linear
size, LS $\leq$15 - 20 kpc), 
powerful ($P_{\rm 1.4 GHz} > 10^{25}$ W Hz$^{-1}$) extragalactic
objects generally associated with
distant (z$>$0.2) galaxies and quasars. 
Their main characteristic is the steep radio spectrum 
($\alpha>$ 0.5, $S\propto \nu^{-\alpha}$) in the optically thin regime,
that flattens and turns over at low frequencies, between
1 GHz and 30 MHz \citep{odea98}. The genuine youth of 
these objects
was strongly supported by the determination of both the kinematic
\citep[e.g.][]{owsianik98,polatidis03,polatidis09,giroletti09,mo10}
and radiative 
\citep[e.g.][]{murgia99,murgia03,mo07} ages of a dozen of the most compact
objects (LS $\leq$ 100 pc) which turned out to be about
10$^{3}$--10$^{5}$ years. Given their young age and the intrinsically
small size,
CSS/GPS sources provide us
with a unique opportunity to study how the radio emission evolves and
which role the ambient medium plays on their growth during the first
stages of their evolution. A dense,
  inhomogeneous environment may slow down the expansion of the radio
  source in the case one jet interacts with a dense cloud. Although
  the confinement may not last for the entire source lifetime, it may
  cause an underestimate of the source age.\\
The radio morphology of CSS/GPS radio sources closely resembles
that of Fanaroff-Riley type-II radio galaxies \citep{fr74},
but on much smaller scale. 
Their radio  structure 
is usually termed {\it symmetric} in the sense that the radio emission is
found on the two opposite sides of the core 
(when detected), giving rise to the classification of either compact
symmetric objects \citep[CSOs,][]{wilkinson94}, if they are smaller
than 1 kpc, or medium-sized symmetric objects \citep[MSO,][]{fanti95}
if they extend on scales up to 15-20 kpc. However, a large fraction of
CSS and GPS sources have a very asymmetric two-sided morphology
\citep[e.g.][]{saikia03} where
one of the lobes is much brighter and closer to the core than the
other. This kind of asymmetry cannot be explained in terms of beaming
effects and path delay, suggesting that the two jets are piercing
their way through an inhomogeneous medium
\citep[e.g.][]{mo07,jeyakumar05}. A strong indication of asymmetries
produced by a jet-cloud interaction comes from the detection of
atomic hydrogen in absorption only against the brighter (and closer to core) 
lobe in the CSS objects 3C\,49 and 3C\,268.3
\citep{labiano06}, and in the restarted source 3C\,236
\citep{conway99}. 
Moreover, evidence that the ionized gas may be
inhomogeneously distributed around the radio source was provided by
the detection of asymmetric free-free absorption against the two lobes
in three of the most compact radio galaxies (LS $<$ 16 pc): 
J0428+3259, J1511+0518 \citep{mo08}, and OQ\,208
\citep{kameno00}. 
All these indications suggest that in a substantial fraction
of small objects the radio source is not uniformly enshrouded by a homogeneous
environment and the two jets may experience different
conditions during their propagation through the interstellar
medium. Although the gas is not dense enough to frustrate the jet
expansion for the entire source lifetime
\citep[e.g.][]{aneta05,fanti00,fanti95}, it may slow down the
source growth if a jet-cloud interaction takes place. The high
fraction of asymmetric intrinsically-compact radio galaxies strongly
supports this scenario. \\
In this paper we present results of global-VLBI and VLA observations at 5 GHz
aimed at determining the radio morphology and the polarimetric properties
of 10 out of the 16 sources of the CSS sample selected by \citet{dd95}
from the Peacock and Wall catalogue \citep{pw81}. Furthermore, the
combination of these data with the 1.7 GHz images already published 
by \citet{dd95} allows the analysis of the spectral index distribution
across the source, which is crucial in order to unambiguously constrain
the nature of each source component.\\
The paper is organized as follows: Section 2 describes the radio
data and the data reduction; in Section 3 we report the results and 
a description of each radio source, while discussion and
summary are presented in Sections 4 and 5, respectively.\\
Throughout this paper, we assume 
$H_{0} = 71$ km s$^{-1}$ Mpc$^{-1}$, $\Omega_{\rm M} = 0.27$,
$\Omega_{\Lambda} = 0.73$, in a flat Universe. The spectral index is
defined as $S$($\nu$) $\propto \nu^{- \alpha}$. \\ 

\section{Observations and Data Reduction}

\subsection{The VLBI data}

Global VLBI observations at 5 GHz (6 cm)
were carried out on September 16 and 17, 1991 
for a total of 30 hours, using the
Mark II recording system \citep{clark73} with 1.8 MHz bandwidth in
left circular polarization (LCP), at the stations reported in
Table \ref{station}. 
The data were correlated with the JPL--Caltech VLBI
Correlator at the California Institute of Technology in Pasadena. \\
Each source was observed in snap shot mode 
for about 1.0--1.5 hour of global time plus
1.5--2 hour of each subnetwork (stations in Europe and US were
observing two different sources). For about 24 hours the
VLA (in A configuration) was used in phased--array mode as part of the
network, greatly enhancing the
sensitivity of the observations. This allowed us to have also
simultaneous conventional interferometric data (see Section 2.2).
In addition, the MERLIN array was
jointly observing, but various failures allowed us to use small parts of the
VLBI data taken at Cambridge and Knockin only. 

The correlated VLBI data were read into the Astronomical
Image Processing System (\texttt{AIPS}) package, developed at the National
Radio Astronomy 
Observatory (NRAO) where fringe fitting, amplitude calibration,
editing, self-calibration and imaging were performed. The flux
density scale was calibrated following \citet{cohen75}, using
B0016+731, B0235+164 and B0851+202 (OJ287) as calibration sources, with
adopted flux densities of 1.70, 1.74 and 3.37 Jy respectively, as
derived from the VLA simultaneous observation (see Sect. 2.2). \\
Our earlier 1.7-GHz images \citep{dd95} 
were used as reference in order to solve for phase
ambiguities and to constrain initial self-calibration. 
The standard techniques of editing, mapping and
self--calibration were applied.  Gain self--calibration was
applied only once at the end of the process, adopting a solution
interval longer than the scan length (30 min), in order to remove
residual systematic errors and to fine tune the flux density scale.
The gain corrections obtained from self--calibration were in general
within 2\%. The 1$\sigma$ noise level on the images, measured
far from the source, is in the range of 0.2 -- 0.9 mJy/beam, often not
far from the thermal noise ($\sim$0.17 mJy/beam for a typical
observation). The dynamic range, defined as the peak--brightness over
r.m.s noise level
($S_{\rm p}$/1$\sigma$), varies from $\sim$200 to $\sim$5000. The
resolution achieved is about 2 milliarcseconds.\\
 
\begin{figure*}
\begin{center}
\includegraphics{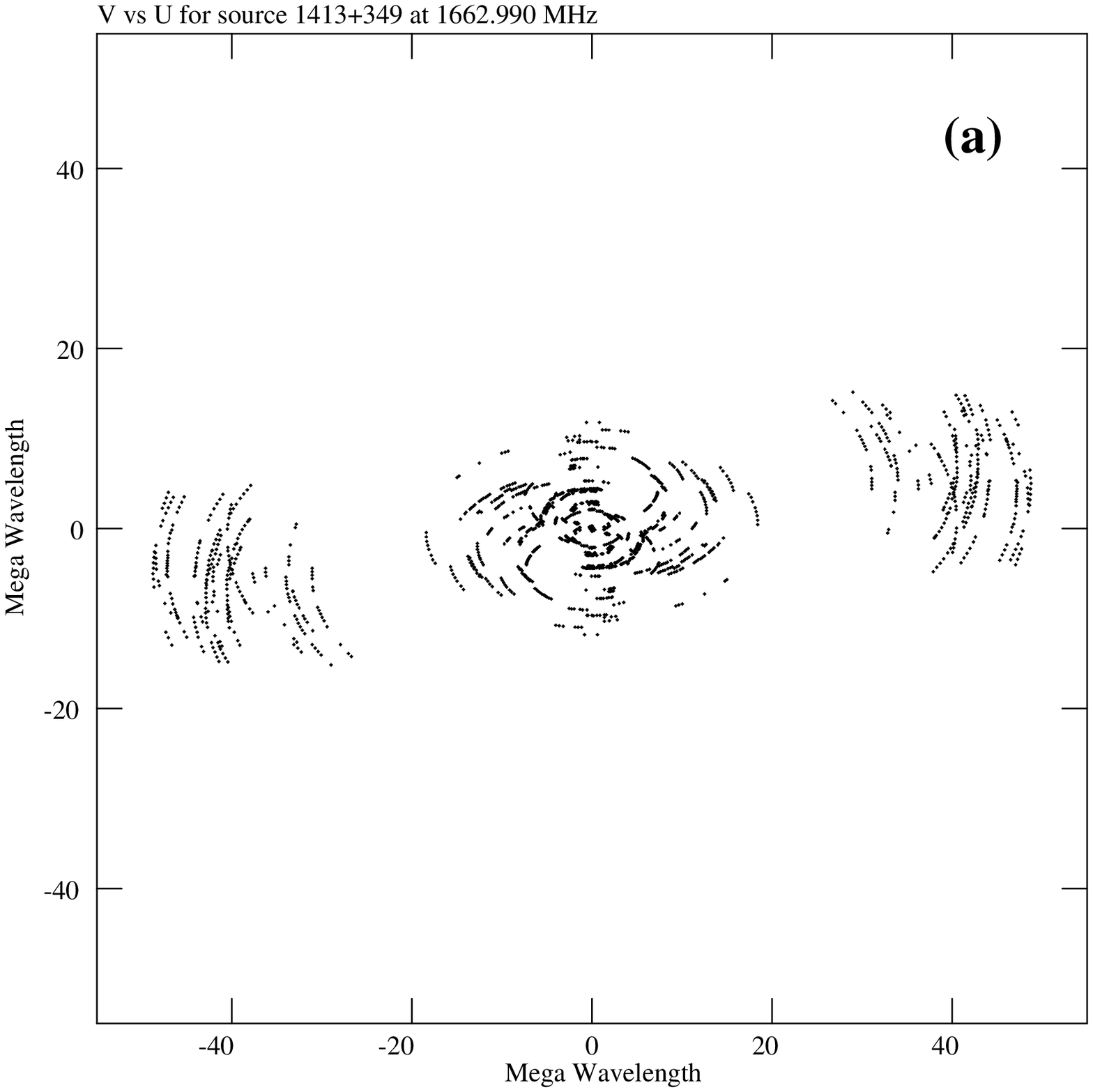}
\includegraphics{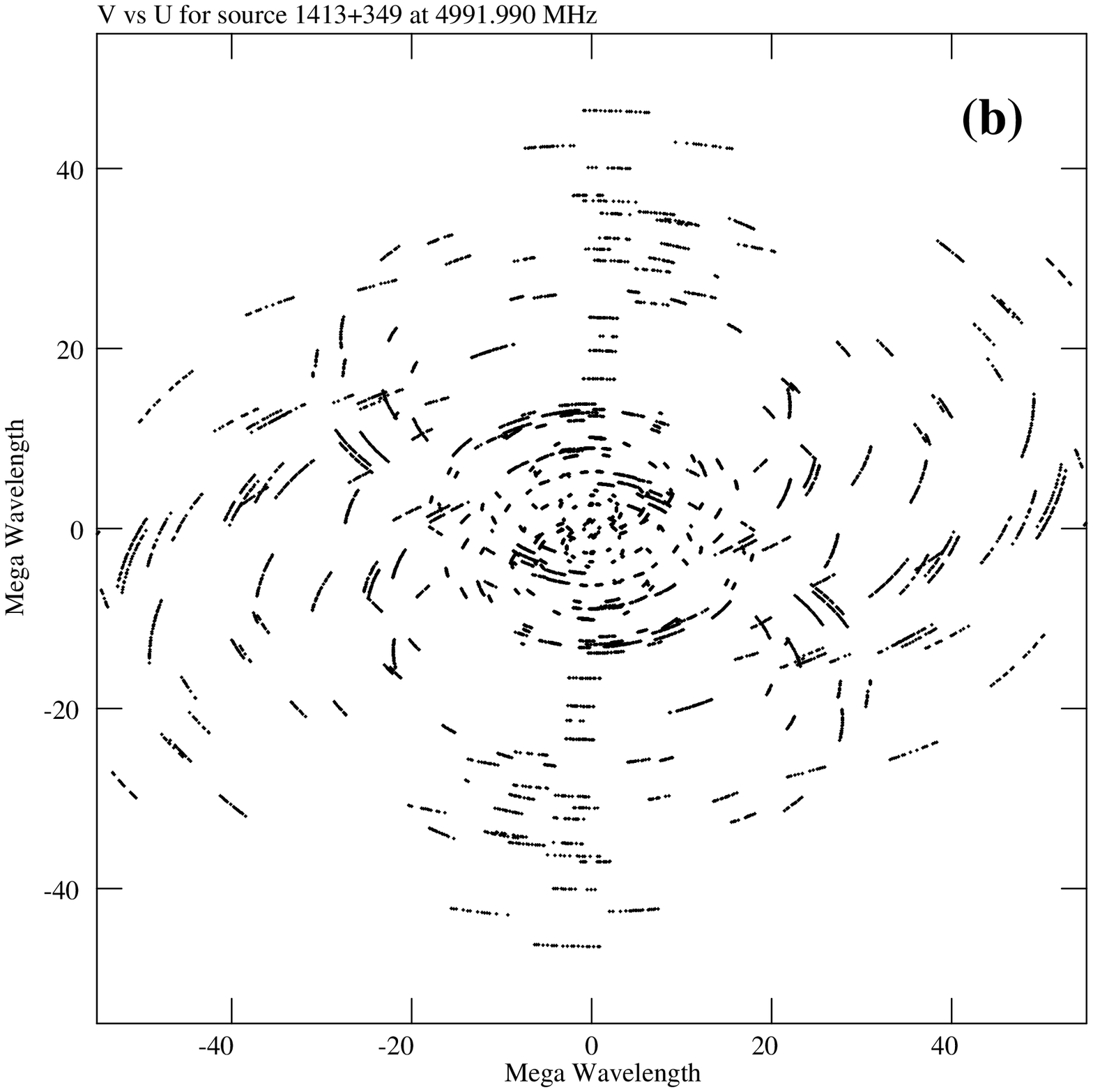}
\vspace{8.7cm}
\caption{{\it uv}-coverage of 1413+349 at 1.7 GHz {\bf (a)} and at
  5 GHz
{\bf (b)}. The latter is restricted to the central region to match
the area covered at 1.7 GHz. Sources at higher declination have better
matches.} 
\label{figuvcov}
\end{center}
\end{figure*}

\begin{table}
\caption{Observational data. 
The diameters listed for the WSRT and the VLA (in brackets) are
equivalent diameters, i.e. 3 antennas for WSRT and 27 antennas for the
VLA. The sensitivities of the VLA refer to a single antenna (0.12)
and to the whole array (2.3).} 
\begin{flushleft}
\begin{tabular}{lccl}
\noalign{\smallskip}
\hline
\noalign{\smallskip}
 Antenna & Diam.(m) & T$_{sys}$ (K) & Sens. (K/Jy) \\
\noalign{\smallskip}
\hline
\noalign{\smallskip}
 WSRT (NL)         &     43   &    &  \\
 Jodrell Bank (UK) &     25   & 38 & 0.12 \\
 Knockin (UK)      &     25   & 38 & 0.1  \\
 Cambridge (UK)    &     32   & 38 & 0.23 \\
 Onsala (S)        &     26   & 50 & 0.06 \\
 Effelsberg (D)    &    100   & 75 & 1.45 \\
 Noto (I)          &     32   & 30 & 0.16 \\
 Haystack (US)     &     37   & 75 & 0.165 \\
 Green Bank (US)   &     43   & 30 & 1.93 \\
 OVRO (US)         &          &    & 0.21 \\
 VLA (US)          & 25 (130) & 35 & 0.12 (2.3) \\
 VLBA$\_$NL (US)   &     25   & 40 & 0.12 \\
 VLBA$\_$FD (US)   &     25   & 40 & 0.12 \\
 VLBA$\_$PT (US)   &     25   & 40 & 0.12 \\
 VLBA$\_$KP (US)   &     25   & 40 & 0.12 \\
 VLBA$\_$LA (US)   &     25   & 40 & 0.12 \\
 \noalign{\smallskip}
\hline
\noalign{\smallskip}
\end{tabular}
\end{flushleft}
\label{station}
\end{table}

\begin{figure*}
\begin{center}
\includegraphics{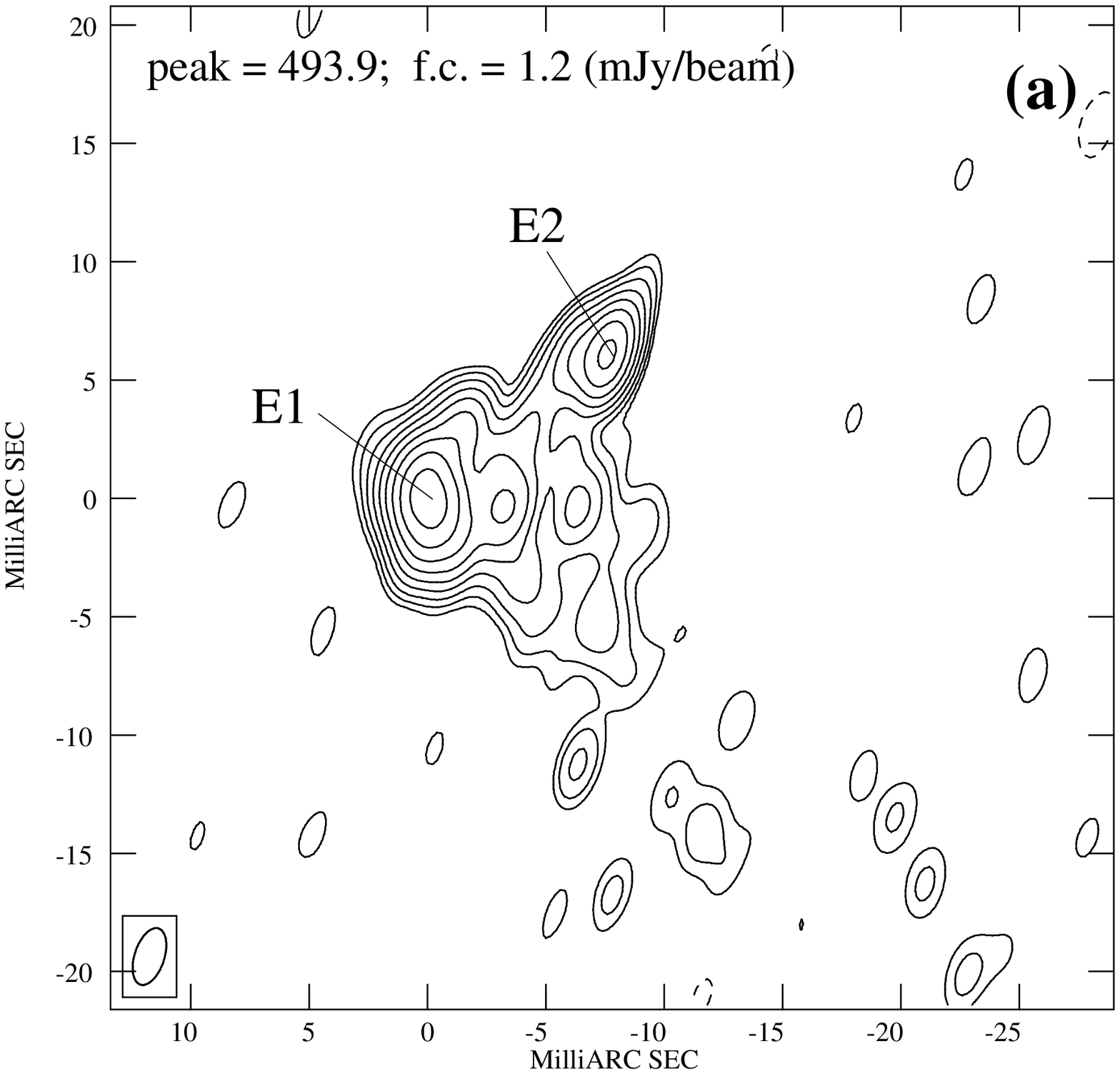}
\includegraphics{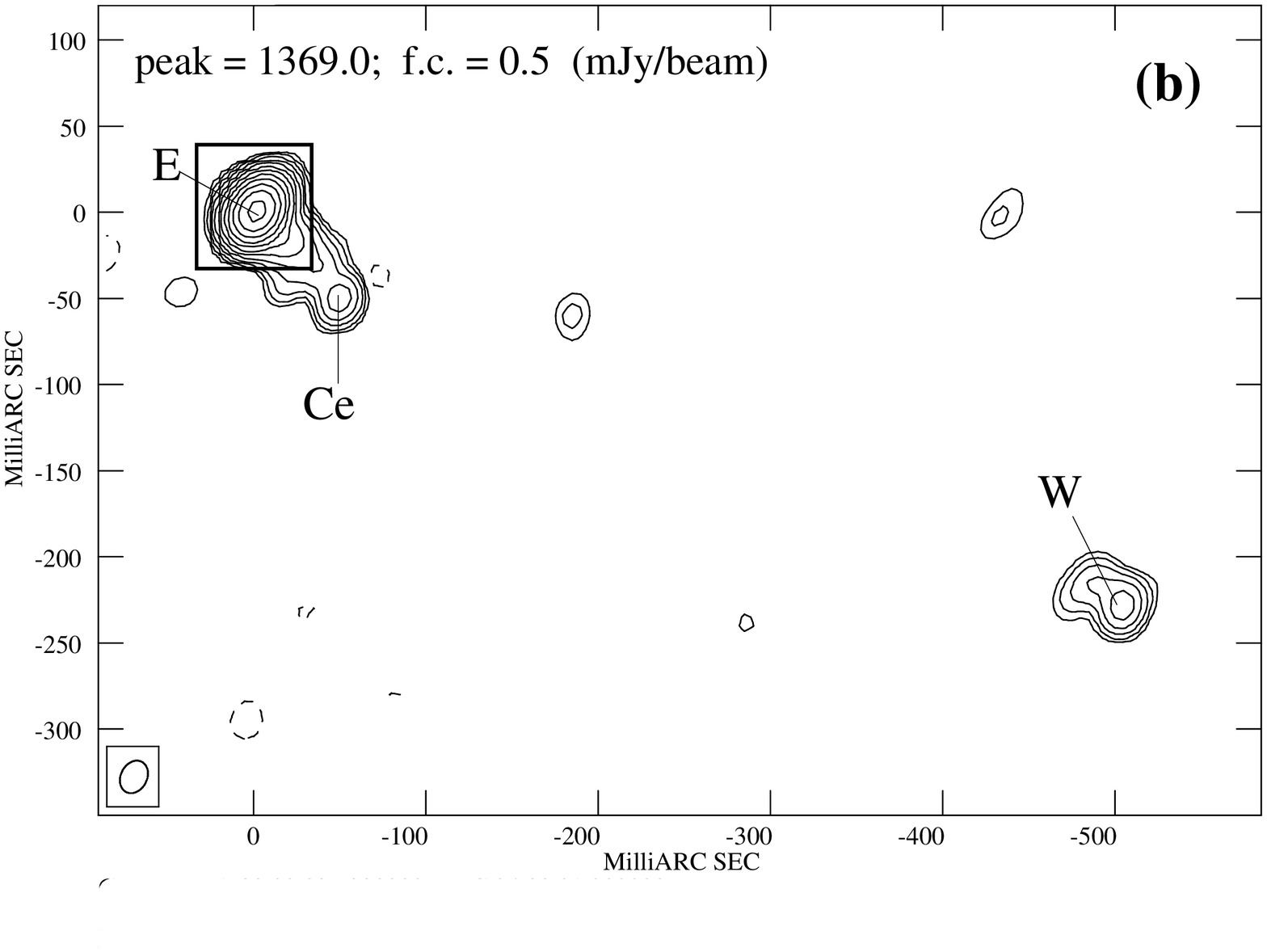}
\vspace{7cm}
\caption{{\bf 0223+341:} Global VLBI images at 5 GHz of the eastern component {\bf
    (a)} and of the total source structure {\bf (b)}. 
The restoring beam is 2.5$\times$1.3 mas$^{2}$ in
p.a. $-$18$^{\circ}$, and 20$\times$15 mas$^{2}$ in
p.a. $-$27$^{\circ}$, respectively, and they are plotted in the bottom
left corner of the image.
In addition, on each image we provide 
the peak flux density in mJy/beam and the first
contour intensity (f.c.) in mJy/beam, which corresponds to three times the off-source
noise level. Contour levels increase by a
factor of 2. The box on panel b represents the area shown with higher resolution
in panel {\bf (a)}.}
\label{0223}
\end{center}
\end{figure*}

\begin{figure}
\begin{center}
\includegraphics{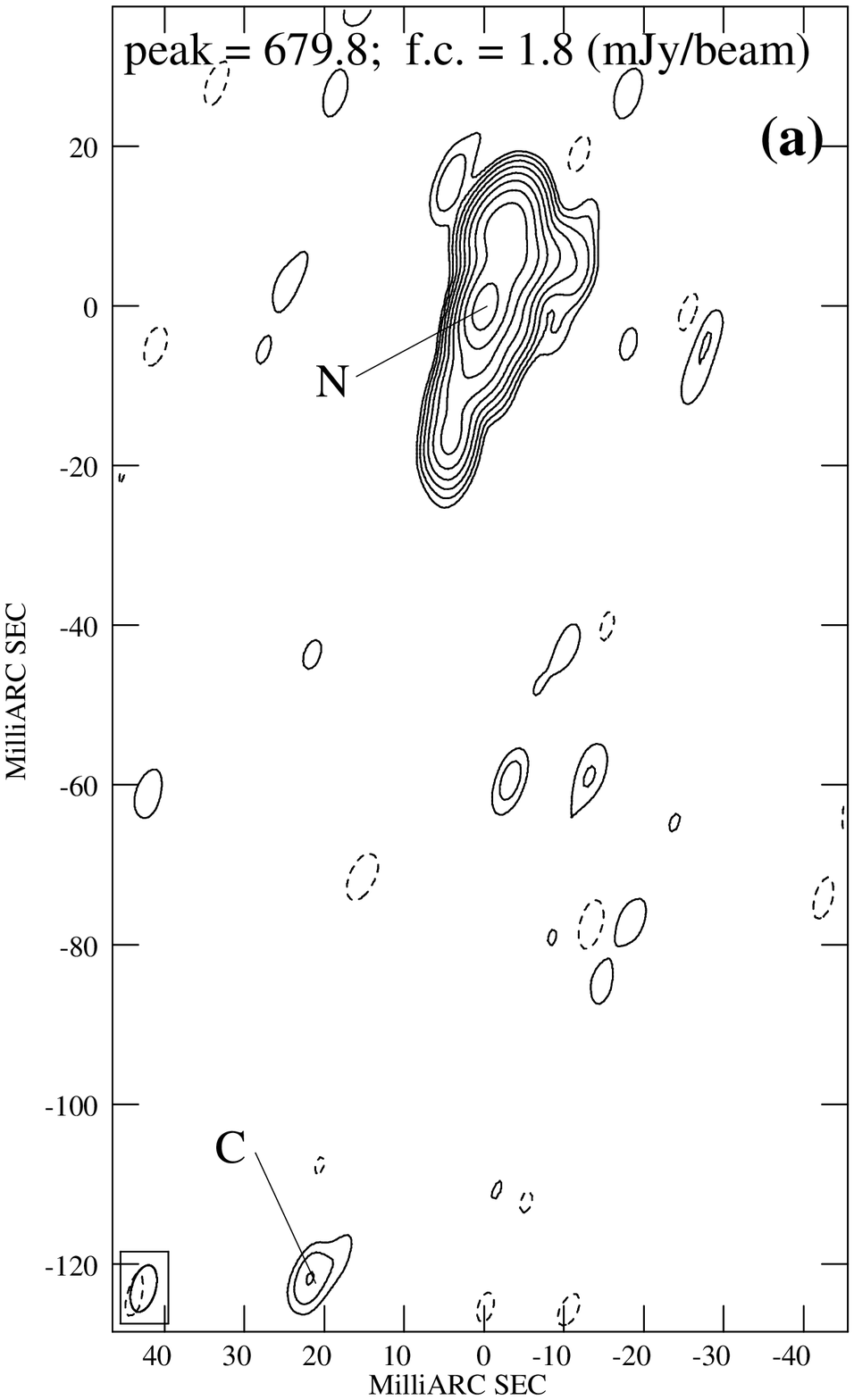}
\includegraphics{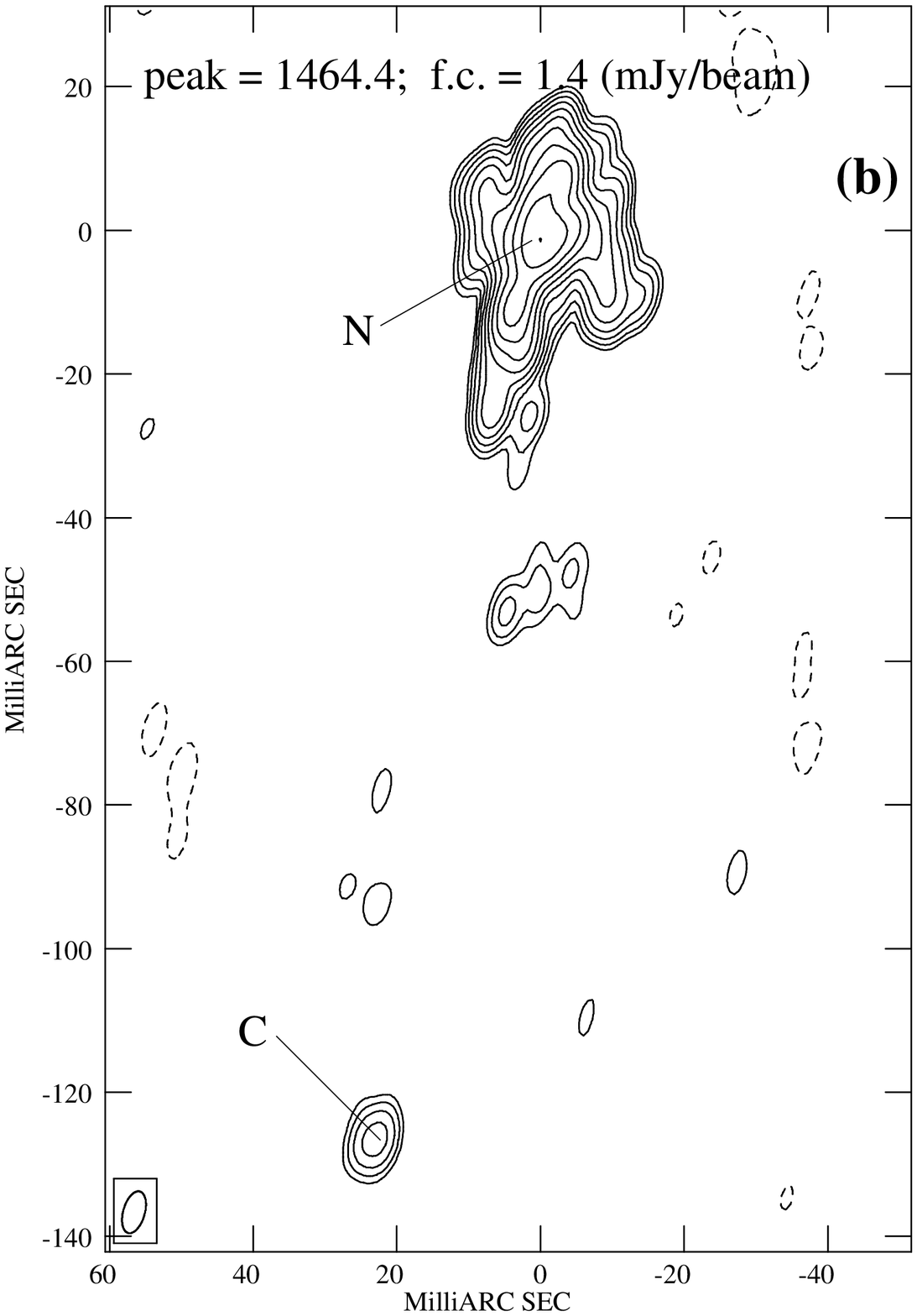}
\vspace{7.5cm}
\caption{{\bf 0316+161:} images of the northern and central components
at 5 GHz {\bf (a)} and at 1.7 GHz {\bf (b)} restored with
an elliptical Gaussian with FWHM = 6 $\times$ 3 mas$^{2}$ in
p.a. $-$16$^\circ$, plotted in the bottom left part of the images. On
each image we provide the peak flux density in mJy/beam, and the first
contour (f.c.) intensity (mJy/beam), which is three times the
off-source noise level. Contour levels increase by a factor of 2.}  
\label{0316}
\end{center}
\end{figure}

\begin{table*}
\caption{Radio properties of the 10 CSS sources from the catalogue
  presented in \citet{pw81} 
and discussed in this paper. Column 1: source name (B1950);
  Col. 2: optical identification; Col. 3: optical
magnitude [inhomogeneous bands]; Col. 4: redshift; Col. 5: linear
scale factor. For consistency with \citet{dd95}, a value of
$z=1$ has been assumed for the sources without redshift; Col. 6: maximum
angular size from the 5-GHz VLBI image; Col. 7: maximum linear size; 
Col. 8: flux density at 5 GHz,
as measured by the VLA during the VLBI observation; Col. 9: flux
density at 5 GHz from the VLBI images;
Col. 10: fractional polarization as measured by the VLA at 5 GHz; 
Col. 11: spectral index between 1.7 and
5 GHz computed using these VLA data and the 1.7-GHz VLA 
total flux density from \citet{dd95}; Col. 12: estimated turnover
frequency \citep{dd95}; Col. 13: references for the optical
  magnitude: 1: \citet{peacock81}; 2: \citet{labiano07}; 3:
  \citet{stickel94}; 4: Sloan Digital Sky Survey  Data Release 9 (SDSS
  DR9) \citep{ahn12}; 5: \citet{stickel96}. } 
\begin{flushleft}
\begin{tabular}{lcllccrccccrcc}
\hline
\noalign{\smallskip}
Source & Opt. Id. & ~m & ~~z & scale & $ \theta_{max}$ & LS &
$S_{\rm VLA, 6}$ & $S_{\rm VLBI, 6}$&
$p_{\rm pol}$ & $\alpha_{\rm thin}$ & $\nu_{\rm m}~ $& Ref.\\
 & & & & pc/mas & arcsec & kpc~~ & Jy & Jy & \% & & MHz & \\
(1)&(2)&(3)&(4)&(5)&(6)&(7)&(8)&(9)&(10)&(11)&(12)&(13)\\
\noalign{\smallskip}
\hline
\noalign{\smallskip}
 0223+341&  Q     &21.3r  &2.910
 &7.898&0.56&4.42&1.75&1.63&$<$0.6 & 0.38 &$<$100& 1\\
 0316+161&  G     &23.4V  &0.907
 &7.830&0.22&1.72&2.89&2.51&$<$0.2 & 0.81 &   800& 2\\
 0319+121&  Q     &18.0V  &2.662 &8.079&0.04&0.32&1.37&1.33&
 5.5  & 0.17 &$<$100& 3\\
 0404+768&  G     &22.0r  &0.598
 &6.662&0.13&0.87&2.95&2.64&$<$0.6 & 0.48 &   350& 1\\
 0428+205&  G     & 18.0r  &0.219
 &3.507&0.10&0.36&2.38&2.00&$<$0.2 & 0.38 &  1150& 4\\
 1225+368&  Q     & 21.4r &1.973
 &8.486&0.06&0.51&0.76&0.73&$<$0.5  & 0.88 & 1750& 4\\
 1358+624&  G     & 19.8r &0.431
 &5.590&0.05&0.28&1.77&1.67&$<$0.5 & 0.70 &  550& 4\\
 1413+349&  EF    &  -    &   -    &8.041&0.03&0.24&0.92&0.87&$<$0.4 &
 0.53 &   850& 4\\
 1600+335&  G     & 23.8R & - &8.041&0.07&
   0.42&2.58&2.55&$<$0.5 & 0.15 & 2400& 5\\
 2342+821&  Q     & 20.2R &0.735
 &7.285&0.18&1.31&1.31&1.29&$<$0.7 & 0.83 &   500& 5\\
\noalign{\smallskip}
\hline 
\noalign{\smallskip}
\end{tabular}
\end{flushleft}
\label{vla_parameter}
\end{table*}

\subsection{The VLA A-configuration Data}

During the VLBI observation, the VLA in A-configuration was observing in
phased array mode for 24 hours at 5 GHz, allowing us to obtain also images
with a resolution of about 0.4 arcsecond for all the sources. In this
way we got accurate flux density measurements of each source and
calibrators and we could search for extended radio emission with
angular size up to 10 arcsec. Furthermore, 
we obtained VLA polarization information
for all the sources. The calibration of the instrumental polarization
could not be very accurate since no suitable D-terms calibration
source was observed on a wide range of
parallactic angles and we had to use three different sources covering
together about 150$^\circ$. 
The absolute flux density scale was set on
3C\,286, observed for a few minutes, for which we assumed a flux density
of 7.38 Jy at 5.0 GHz. 

\section{Results}

In this section we present the total intensity VLBI images at 5 GHz for
the 10 sources discussed in this paper and listed in Table
\ref{vla_parameter}, together with spectral index images
obtained by comparing these 5-GHz data with those at 1.7
  GHz presented in 
\citet{dd95}. For each source we provide a description of the main
characteristics.\\  

\subsection{Source images}

In general, for each source we obtained images with a set of different
restoring beams and samplings in order either to match the resolution of our
1.7-GHz images, or to get all the details of the structures. \\
The $uv$-coverages at 1.7 and 5 GHz for the source 1413+349 are shown in
Fig. \ref{figuvcov} as an example: the 5-GHz visibilities are
restricted to the 
region covered by the 1.7-GHz data. Sources with higher declinations
have a better match between the 1.7-GHz and 5-GHz $uv$-coverages.\\
High resolution images at 5 GHz of each source are presented in
Figs. \ref{0223} -- \ref{2342}. For the source 0316+161 we present
also the
high resolution image at 1.7 GHz with the identification of the source
core, which was not provided in the paper by \citet{dd95}. 
In the case of 0223+341, where one lobe is located at about 0.5 arcsec
from the main component, and  
1600+335, where the
central component is surrounded by 
an extended low surface brightness
feature, a lower resolution
image is presented next to the
high-resolution one. \\
Source parameters (total flux density and deconvolved size) 
have been generally measured by means of the \texttt{AIPS} task JMFIT 
for marginally resolved components. When a component could not be properly 
fitted with a Gaussian profile, TVSTAT was used to measure the flux
density, while 
the angular size was measured from the lowest contours and it
corresponds to 1.8 times 
the size of the full width at half maximum (FWHM) 
of a conventional Gaussian covering a similar area 
\citep{readhead94}.  \\
Sub-components are referred to as North (N), South (S), East (E), West
(W), Central (Ce), Jet (J), and flat spectrum core (C) when detected,
following the labelling used in \citet{dd95}.
The parameters of the sub-components 
are given in Table \ref{component}. \\
To produce the spectral index distribution across each source, 
we created low-resolution images
with the same restoring beam, {\it uv}-range and image sampling used for the
1.7-GHz image. Images at 1.7 and 5 GHz 
were obtained using all the baselines
shorter than 50 M$\lambda$. These images were then convolved with a larger
beam (usually between 6 and 10 mas), in order to highlight features
corresponding to baseline-length shorter than 20 M$\lambda$, where the
mismatch of the {\it uv}-coverages is less significant.
We must note that the addition of the VLA array in the 5 GHz observations
reduces the gap at the short spacings, providing a good set of short
baselines that allows a good sensitivity to extended, low-surface
brightness structures. \\
The spectral
index images were produced with the \texttt{AIPS} task COMB. 
Blanking was done
considering only the pixels of the input images with values above five times
the rms measured on the off-source image plane. Image registration
was performed using the task LGEOM in \texttt{AIPS}, 
by comparing the location of optically-thin, compact bright features. 
This led to satisfactory results in case of compact regions, while
artificial gradients may be present in the extended structures,
precluding an accurate analysis of these regions.
Furthermore, we do not provide spectral index images
of the more extended regions, like the relatively large and low surface
brightness lobes in 0223+341, 0316+162, and 0428+205, given that most
of the information in the 1.7-GHz images relies on the MERLIN baselines
\citep{dd95}, which are not sampled by the 5-GHz VLBI observations.\\
In Fig. \ref{spix}, for all the sources we show the greyscale 
spectral index images between 1.7 and 5 GHz superimposed on the
5-GHz contours obtained from the low resolution images.\\

No VLA A-configuration 
images are presented since the resolution of about 0.4 arcsec
is not enough to resolve the
targets, and only the largest object in our sample, 0319+121,
is marginally resolved.\\

\subsection{Notes on individual sources}

Here we provide a description of the radio morphology of the
observed sources. In addition to the {\it local} spectral index
derived on the spectral index images (Fig. \ref{spix}), 
we computed the {\it total}
spectral index integrated on the whole component by means of the total
flux density measured on the high-resolution images at 5 and 1.7 GHz.
The formal error on both the {\it total} and {\it local} 
spectral index is about 0.1, and it
has been calculated assuming the error propagation theory.
However, we note that the use of the total flux density in
computing the spectral index may cause an artificial steepening due to
the absence of the shortest spacing at the higher frequency, implying
that the formal errors associated to the extended components may be a
lower limit. The {\it total} spectral index values are
reported in Table \ref{component}.\\

\begin{table}
\caption{Size and flux density of the components in the VLBI images.
Columns 1 and 2: source name and component label; Cols. 
3, 4, and 5: deconvolved major and minor axes, and the position angle of the
major axis derived from the fit to the image. 
The values reported are
the FWHM of the fitted Gaussian component; Col. 6: 
VLBI flux density at 5 GHz; Col. 7: spectral index computed from the
full-resolution images at 1.7 and 5 GHz; 
Col. 8: the equipartition magnetic
fields (see Section 4.3). For consistency with \citet{dd95},
when the redshift is not available we assume $z=1$ 
in the determination of the magnetic field and the value is
  reported in italics. }
\begin{center}
\begin{tabular}{llrrrrrr}
\hline
Source & Comp. & $\theta_1 $ & $\theta_2 $ &pa & $S_{5}$&$\alpha_{1.7}^{5.0}$ &$H_{\rm eq}$\\
       &       & mas & mas & $^{\circ}$ & mJy & &mG\\ 
(1)&(2)&(3)&(4)&(5)&(6)&(7)&(8)\\
\hline 
0223+341     & E1  &  2.2&  0.8&  24 &  872 & 0.5&66 \\ 
             & E2  &  1.1&  0.6& 117 &  239 & 0.0&66 \\ 
          & E\tablenotemark{a}&   7.6&  3.2& 127&  1545 &0.3&25 \\ 
          &W\tablenotemark{a}   & 21.4&  5.7&  48 &   20 &2.5 &4 \\ 
0316+161     & N\tablenotemark{b}&48.5 & 24.6 &165& 2560 &0.8 &4 \\ 
             & C  &    2.9 &  2.4 &168&   18  &0.3 &5 \\ 
             & S\tablenotemark{a}&37.6 &  9.5 &131& 33   &2.4 &1 \\ 
0319+121  & C\tablenotemark{c}& 1.3 & 0.2 & 173 & 824   &0.2 &121 \\ 
          &J1     & 2.0 & 1.4 & 3 & 219      &    &31 \\ 
          &J\tablenotemark{b} &38.5& 4.3 & 170 & 508    & 0.7 &15 \\ 
0404+768  &W    &   7.8& 5.7 & 147 &  1100   &   &7 \\ 
          &W\tablenotemark{b}& 55.8&31.6 & 105 & 2304   &0.5&3 \\
          &C    &  4.1 & 0.9 & 7  & 182     &-0.5&14 \\ 
          &E\tablenotemark{b}& 21.0 & 21.0 & - & 157 &0.9& 2 \\ 
0428+205 & C\tablenotemark{c}& 1.0 & 0.2 & 156 & 29 &-0.4& 25 \\ 
         & J     & 3.6 & 0.5 & 162 & 135 &  &16 \\ 
         & S     & 2.3 & 1.6 & 129 & 538 &  &14 \\ 
         & S\tablenotemark{b}&23.2 & 9.0 & 135 &1777 &0.4&6  \\
1225+368 & E1    & 2.9 & 1.3 & 142 & 457 &0.8& 28 \\ 
         & E2    & 1.5 & 0.9 & 64  & 91  &1.0&27 \\
         & E3    & 2.1 & 1.7 &  -  & 41  &1.0&13 \\ 
         & C     & 2.2 & 2.2 &  -  & 62  &0.3&13 \\ 
         & W\tablenotemark{b}& 4.5 & 4.5 &  -  & 29  &1.4& 10 \\ 
1358+624 & N\tablenotemark{b}&15.0 & 10.0 & 120 & 391 &1.0& 5 \\ 
         & C     & 2.3  & 1.1 & 116 & 38  &-0.3&9 \\
         & S\tablenotemark{b}&33.5 & 11.5 & 120 & 1250 &0.6&5 \\ 
1413+349 & C     & 3.5 & 0.7 & 32 & 607   &0.3&{\it 28} \\ 
         & E\tablenotemark{b}&23.3 & 7.0 & 40 & 227    &0.7& {\it 5} \\ 
1600+335 & C\tablenotemark{c}& 1.6 & 0.4 & 17 & 1926  &0.0&{\it 67} \\ 
         & J     & 3.0 & 1.5 & 173 & 351  &  &{\it 16} \\ 
         & E\tablenotemark{a,b}&35.5&18.1 & 100 & 97  &0.5& {\it  2} \\ 
2342+821 & W\tablenotemark{b}& 27.8 & 22.2 & 85 & 1181 &0.8& 4 \\ 
         &Ce\tablenotemark{b} & 16.7 & 5.5 & 130 & 59  &1.0& 5 \\ 
         &E\tablenotemark{b} & 7.7   & 3.3 & -   &  25 &1.0& 5 \\ 
\hline
\end{tabular}
\end{center}
\tablenotetext{a}{The component parameters have been derived on the
low resolution map.} 
\tablenotetext{b}{The component angular size has been measured on the contour
image, while the flux density has been derived by means of TVSTAT.}
\tablenotetext{c}{The spectral index refers to the combination of
components C and J (J1 in the case of 0319+121). }
\label{component}
\end{table}

\subsubsection{0223+341~~[~Q, ~$m_r$=21.3, ~z=2.91]} 

The radio source 0223+341 (alias 4C\,34.07) 
is optically identified with a quasar at redshift
$z=2.91$ \citep{willott98}. 
Our global VLBI image at 5 GHz (Fig. \ref{0223}a) 
has enough resolution to
reveal the substructure of the most compact region labelled East in the
image published at 1.7 GHz \citep{dd95}.
This component has a very complex morphology dominated 
by two resolved bright knots separated by $\sim$ 10 mas (79 pc), E1 and E2,
located at the northern and eastern tips of the structure. In the SW
direction the source is resolved in several blobs pointing towards
the western lobe,
visible in MERLIN images \citep{dd95}. In our 5-GHz
observations the western lobe, W, is detected at a few mJy level only 
(Fig. \ref{0223}b) implying a steep
spectrum, 
although the lack of short spacings may have
prevented the detection of the large scale 
diffuse emission. The central compact component, Ce,
accounts for 16 mJy, and is located $\sim$68 mas (537
pc) from component E at a position angle (p.a.) of
43$^{\circ}$. On the other hand, component W is at about 490 mas (3.8
kpc) from component Ce, with a position angle of -110$^{\circ}$. 
Since the compact component Ce is completely
resolved in the high-resolution image, its interpretation as the
source core is unlikely.\\
Although the {\it total} spectral index of component E is $\alpha = 0.3$ (Table \ref{component}), 
the analysis of the {\it local} spectral index distribution (Fig. \ref{spix}a) 
indicates different values for E1 and E2: $\alpha \sim 0.3$, and
$\alpha \sim 0$, respectively. This difference may be an effect of 
different physical sizes in presence of synchrotron self-absorption
(SSA), which is more effective on the smaller
component E2. A possible
interpretation of the flat spectra and the complex
structure, reminiscent of
the hotspot 3C\,20 East \citep{hardcastle97} although on much smaller
scales, is that  
knots E1 and E2 may be the primary and secondary components of 
a double hotspot. In this scenario the source core may be
  self-absorbed even at 5 GHz, and is located
  between component E and W.
On the other hand, VLBI observations at 327 MHz (Lenc et
  al. 2008, Dallacasa et  al., in preparation), pointed out the
  presence of a northern component located at about 0.24 arcsec ($\sim$2.4
  kpc) from component E, 
  which was not detected in the MERLIN-VLBI images presented by
  \citet{dd95}. The presence of the northern component may indicate
  that component E is likely hosting the source core. The spectral index
  computed between 327 MHz and 1.7 GHz turns out to be flat ($\alpha
  \sim 0.2$) supporting this interpretation. \\
Almost 96\% of the total flux density measured by the VLA is recovered
by our global-VLBI observations.\\ 
No significant polarized emission has been detected by the VLA at 5
GHz ($p <0.6\%$).\\

\subsubsection{0316+161~~[~G, ~$m_v$=23.4, ~z=0.907]}

The radio source 0316+161 (alias CTA\,21)
is optically identified with a galaxy at redshift
$z=0.907$ \citep{labiano07}. 
Our global VLBI image at 5 GHz enables us to identify 
the source core, labelled C in Fig. \ref{0316}, that is located at
about 125 mas ($\sim$980 pc) from component N, 
and about 80 mas ($\sim$626 pc) from component S. 
Its presence was already found in
the 1.7-GHz image \citep{dd95}, but both its location slightly misaligned ($\sim$
25$^{\circ}$) with respect to the two lobes, and
the lack of spectral index information did not allow a secure identification.
At 5 GHz the core flux density is 18 mJy, i.e. 0.6\% of the total
flux density measured at the VLA. The radio emission is dominated by
the northern lobe which accounts for almost 90\% of the total source
flux density. Its complex morphology shows an elongated structure in
the north-south direction with a bright and compact region, likely the
hotspot, located
almost at the centre of the structure. A diffuse emission, also seen in
the 1.7-GHz image, is located perpendicularly to the main structure. 
The southern lobe is
visible in the 1.7-GHz image \citep{dd95} as an amorphous region with
no clear indication of the presence of a hotspot, while it is resolved
out by these 5-GHz observations. Therefore, we
assumed that the end of the jet is located at the edge of the lobe in
the 1.7-GHz image.
The
flux-density ratio between the northern and the southern structures
has been estimated on the 1.7-GHz image and is $R_{\rm S} \sim 14$,
while the arm-length ratio is $R_{\rm R} \sim 0.6$, indicating a
brighter-when-closer behaviour. \\ 
Component N has a {\it total} spectral index $\alpha \sim0.8$ over the
whole structure (Table \ref{component}), 
with some small local fluctuations.
The spectral index of the core region is $\alpha \sim 0.3$,
although this value may be contaminated by the presence of the
jet base (Fig. \ref{spix}b). 
The {\it total} spectral index of the southern lobe is
artificially steep ($\alpha=2.4$, Table \ref{component})
due to the flux density of the extended structure
that could not be detected by our 5-GHz VLBI observations. \\
Almost 91\% of the total flux density measured by the VLA is recovered
by our global-VLBI observations, and a substantial fraction of the
missing flux density could be ascribed to the southern lobe.\\
No significant polarized emission has been detected by the VLA at 5 GHz
($p <$ 0.2\%), which is consistent with what was found
by \citet{stanghellini98}. \\

\begin{figure}
\begin{center}
\includegraphics{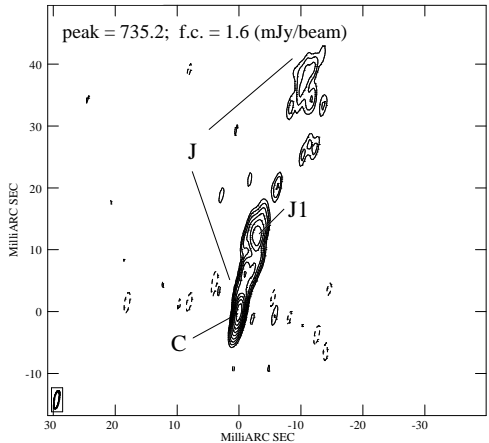}
\vspace{7cm}
\caption{{\bf 0319+121:} global VLBI image at 5 GHz. The
  restoring beam is 3.8$\times$2.5 mas$^{2}$ in p.a. -15$^{\circ}$,
  and is plotted on the bottom left corner. On the image we provide
  the peak flux density in mJy/beam, and the first contour (f.c.)
  intensity (mJy/beam), which is three times the off-source
  noise level. Contour levels increase by a factor of 2.}
\label{0319}
\end{center}
\end{figure}

\subsubsection{0319+121~~[~Q, ~$m_v$=18.0, ~z=2.662]} 

The radio source 0319+121 (alias PKS\,0319+12)
is optically identified with a quasar at
redshift $z=$2.662 \citep{stickel94}. 
The radio emission at 5 GHz is dominated by a
compact component, labelled C in Fig. \ref{0319}, close to the
southernmost edge of the structure  
accounting for 60\% of the total flux density
measured with the VLA.
A jet structure, J, emerges from this bright component with a position 
angle of $-14^{\circ}$. At about 40 mas ($\sim$320 pc) the jet bends slightly
towards the diffuse arcsecond-scale emission already detected by
VLA observations \citep{murphy93}. Although the {\it total} spectral index
computed on the whole source is rather flat ($\alpha=0.2$, Table
\ref{component}), the analysis of the {\it local} spectral index distribution
(Fig. \ref{spix}c) shows a steepening of the spectrum as we move away
from the flat spectrum ($\alpha \sim 0.0$) core, towards the outermost
region where $\alpha \sim 0.7$.  
Almost 98\% of the total VLA
flux density is recovered by our global VLBI observations.\\
The VLA observation at 5 GHz points out a fractional polarization of 5.5\% 
(total polarized flux density of 75 mJy). The integrated polarization
angle at the observed frequency is $\sim 65^\circ$ and the
inferred magnetic field is roughly parallel to the VLBI jet axis, if we assume
little rotation measure (RM) for this source. \\

\begin{figure}
\begin{center}
\includegraphics{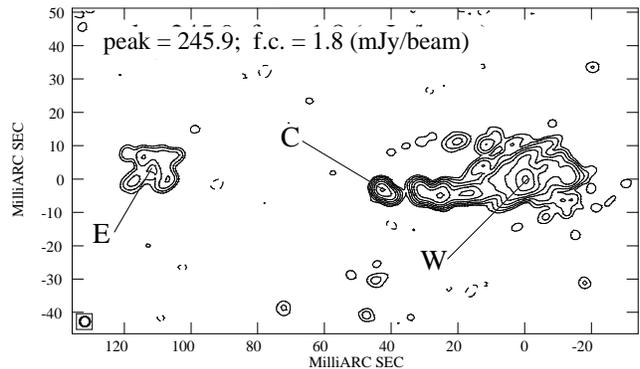}
\vspace{5.5cm}
\caption{{\bf 0404+768:} global VLBI image at 5 GHz. The  
image is restored with a circular Gaussian of
FWHM = 4.0 mas, and is plotted on the bottom left corner. On the image we
provide the peak flux density in mJy/beam, and the first contour
(f.c.) intensity (mJy/beam) which is three times the off-source noise
level. Contour
levels increase by a factor of 2. The image is rotated
counterclockwise by 45$^{\circ}$.} 
\label{0404}
\end{center}
\end{figure}

\subsubsection{0404+768~~[~G, ~$m_r$=22.0, ~z=0.5985]} 

The radio source 0404+768 (alias 4C\,76.03) is optically identified with a
galaxy at $z=0.5985$ \citep{odea91}.
The image at 5 GHz (Fig. \ref{0404}) confirms the morphological interpretation
proposed in \citet{dd95} (in both cases the image is rotated
counterclockwise by 45$^{\circ}$).  
The core region, C, hosting both the source core and the jet base, is
located between the two lobes 
at the easternmost edge of the visible 
jet. The core flux density is 182 mJy, i.e. 6\%
of the total flux density measured by the VLA. The radio structure 
is very asymmetric: the core is located about 70 mas (465 pc) and 40
mas (265 pc) from components E and W, respectively. A
one-sided jet connects the core with the western lobe. The
radio emission is dominated by component W, that
accounts for almost 80\% of the total VLA flux density. The
flux-density ratio between the western and the eastern structures 
is $R_{\rm S} \sim 9.5$ and 7 at 1.7 and 5 GHz, respectively, 
while the arm-length ratio is $R_{\rm R} \sim0.6$,
implying a brighter-when-closer behaviour. \\
Component W has a {\it total} spectral index of
$\alpha = 0.5$ (Table \ref{component}), indicating that the radio
emission is dominated by the hotspot.
The core has an inverted spectrum ($\alpha \sim -0.5$, Table \ref{component}).
The spectral index of component E is
artificially steeper likely due to the emission of the extended structure
that cannot be detected by our 5-GHz VLBI observations (Fig. \ref{spix}d). \\ 
About 90\% of the total VLA flux density could be recovered by our
global-VLBI observations. \\
No significant polarized emission has been detected by the VLA at 5
GHz ($p <$ 0.6\%).\\
Deeper VLBA observations in L and C bands will be presented in a
forthcoming paper which will focus on an accurate spectral analysis of this
source. 

\begin{figure}
\begin{center}
\includegraphics{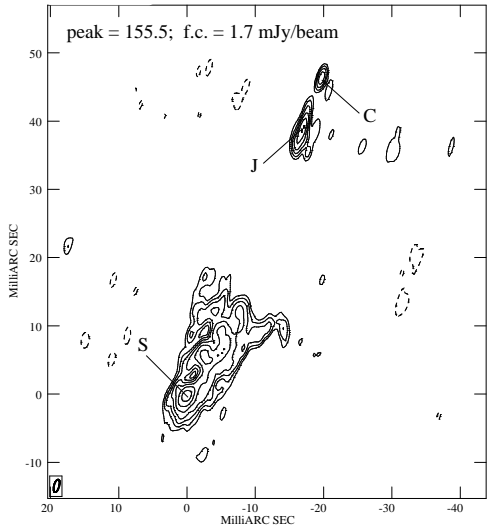}
\vspace{7.5cm}
\caption{{\bf 0428+205:} global VLBI image at 5 GHz. The restoring
  beam is 1.9 $\times$0.75 mas$^{2}$ in p.a. $-15^{\circ}$, 
and is plotted on the bottom left corner. On the image we provide the peak
flux density in mJy/beam, and the first contour (f.c.) intensity
(mJy/beam), which
is three times the off-source noise level. Contour levels increase by a
factor of 2.}
\label{0428}  
\end{center}
\end{figure}

\subsubsection{0428+205~~[~G, ~$m_r$=18.0, ~z=0.219]} 

The radio source 0428+205 (alias DA\,138) is optically identified with a
galaxy at redshift $z=0.219$ \citep{stickel94}. 
Its radio emission at 5 GHz is dominated by the southern and central
components (Fig. \ref{0428}), 
consistent with what was previously found by \citet{fomalont00}.
The northern lobe, located about 175 mas
($\sim$615 pc) from the core, 
is not detected by our 5-GHz observations.
The high-resolution image could resolve the core region into
two main components. From the spectral analysis (Fig. \ref{spix}e), 
we suggest that the core, C, is the northern
component, that accounts for 29 mJy (i.e. 1.2\% of the total VLA flux
density), while the elongated component J 
is likely the main jet. The southern lobe, S is dominated by
a compact component, likely the hotspot, 
located about 50 mas ($\sim$175 pc) from the core.
We estimate the flux-density ratio between the southern and
the northern structure on the basis of the 1.7-GHz
image and we found $R_{\rm S} \sim$ 6, while the arm-length ratio is
$R_{\rm R} \sim 0.3$, implying a brighter-when-closer behaviour.\\
The lower
resolution of the spectral index
image does not allow us to identify the single sub-structures 
(Fig. \ref{spix}e). The
unresolved core region is characterized by a {\it total} spectral
index $\alpha \sim -0.4$. In component S the spectrum
steepens progressively from the southern hotspot, with a {\it local}
spectral index $\alpha \sim
0.4$, towards the core, that is consistent with the dynamic scenario 
in which the electrons deposited towards the centre of the source are older 
than those found closer to the hotspot.\\
In our global-VLBI observations, about 8\% of
the total flux density measured by the VLA could not be recovered, and
it could be ascribed to the northern lobe completely 
resolved out by the present observations. \\
No significant polarized emission has been detected by VLA
observations at 5 GHz ($p <$ 0.2\%).\\ 

\begin{figure}
\begin{center}
\includegraphics{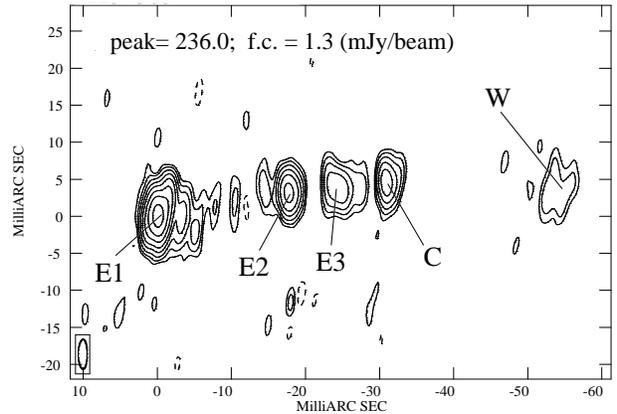}
\vspace{6cm}
\caption{{\bf 1225+368:} global VLBI image at 5 GHz. The
  restoring beam is 4.0 $\times$ 1.2 mas$^{2}$ in p.a. 0$^\circ$, and
  is plotted on the bottom
left corner. On the image we provide the peak flux density in
mJy/beam, and the first contour (f.c.) intensity (mJy/beam), 
which is three times
the off-source noise level. Contour levels increase by a factor of 2.}
\label{1225}
\end{center}
\end{figure}

\subsubsection{1225+368~~[~Q, ~$m_r$=21.4, ~z=1.973]} 

The radio source 1225+368 (ON\,343) is optically identified with a
quasar with a redshift $z=$1.973 \citep{xu94}. This source
has a complex and asymmetric radio morphology 
dominated by a sequence of compact regions aligned in
the EW direction (Fig. \ref{1225}), in good agreement with what was found
by \citet{xu95} at the same frequency.
The availability of data at both 1.7 and 5 GHz allows us to
unambiguously identify the source
core with the flat-spectrum C component ($\alpha \sim 0.3$, Table
\ref{component}). The core flux density is 60 mJy at 5 GHz (i.e. 8\%
of the total VLA flux density at this frequency).  
The radio emission
is dominated by the eastern part of the source, while the western
component, W, represents only 4\% of the total flux density.
The brightest component E1, likely a hotspot, is located at 32 mas (270 pc)
from the core, and shows a {\it local} spectral index $\alpha \sim
0.8$ (Fig. \ref{spix}f). 
The radio spectrum steepens going
backward along the jet components ($\alpha = 1.0$), 
indicating that components E2 and E3, located between the hotspot and the core,
are likely knots in the jet.
Component W is located at 23
mas (200 pc) from the core, and has a very steep
spectrum ($\alpha \sim 1.4$) without any evidence of compact
regions. The hint of the jet visible in the 1.7-GHz image has
disappeared at this higher frequency.\\
The flux density ratio computed between the eastern and western
structures is $R_{\rm
  S} \sim 12.8$ and 20.3 at 1.7 and 5 GHz, respectively, 
while their arm-length ratio is $R_{\rm R} \sim 1.4$,
providing a brighter-when-farther behaviour.\\
The total flux density recovered by our global-VLBI observations
is in agreement with the value found with the
VLA, excluding the presence of significant emission on
scales larger than those represented in our VLBI image.\\ 
No significant polarized emission has been detected by the VLA at 5 GHz
($p <$ 0.5\%).\\

\begin{figure}
\begin{center}
\includegraphics{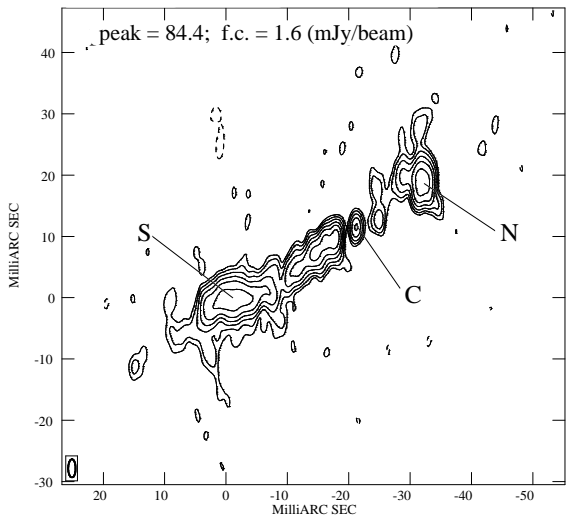}
\vspace{7.5cm}
\caption{{\bf 1358+624:} global VLBI image at 5 GHz. The
  restoring beam is 3.0 $\times$ 1.2 mas$^{2}$ in p.a. 0$^\circ$, and
  is plotted on the bottom
left corner. On the image we provide the peak flux density in
mJy/beam, and the first contour (f.c.) intensity (mJy/beam), which is
three times the off-source noise level. 
Contour levels increase by a factor of 2.}
\label{1358}
\end{center}
\end{figure}

\subsubsection{1358+624~~[~G, ~$m_r$=19.8, ~z=0.431]} 

The radio source 1358+642 (alias 4C\,62.22) is optically identified
with a galaxy at redshift $z=0.431$ \citep{lawrence96}.
In our 5-GHz observations, 
this source has a complex and asymmetric structure that resembles the
sources B3\,1242+410 \citep{mo04}, and 1946+708 \citep{peck01}. The
morphology shown in Fig. \ref{1358} is in good agreement with VLBA
observations presented by \citet{taylor96}. The
radio emission is dominated by the southern jet, S, while the northern
component, N, accounts for 22\% of the total flux density. 
The core region is hosted in component C, which is characterized by an
inverted spectrum ($\alpha \sim -0.3$). 
The core flux density is 38 mJy at 5 GHz, i.e. about 2\%
of total flux density. 
The {\it total} spectral index of component N 
is $\alpha = 1.0$, while some local
flattening can be seen in its NE edge ($\alpha \sim 0.6$,
Fig. \ref{spix}g), at
about 15 mas ($\sim$84 pc) from the core, where a hotspot may be
present.
The southern jet has a spectral index $\alpha
\sim 0.6$. Its structure is well collimated in the initial part, 
and broadens farther out ending in a diffuse, steep-spectrum emission.
The flux-density ratio between the southern
and the northern structure is $R_{\rm S} \sim 2.2$ and 3.2 at 1.7 and
5 GHz, respectively, while the
arm-length ratio is $R_{\rm R} \sim 1.9$, indicating a
brighter-when-farther behaviour.\\
The total flux density derived from our VLBI observations matches
the value measured with the VLA, indicating that no significant 
extended structure
on a larger scale is present. \\
No significant polarized emission has
been detected by the VLA at 5 GHz
($p <$ 0.5\%). However single-dish measurements at 15 GHz \citep{aller85}
find significant fractional polarization ($p \sim$ 5\%) suggesting
that this object may be highly depolarized at lower frequencies.\\

\begin{figure}
\begin{center}
\includegraphics{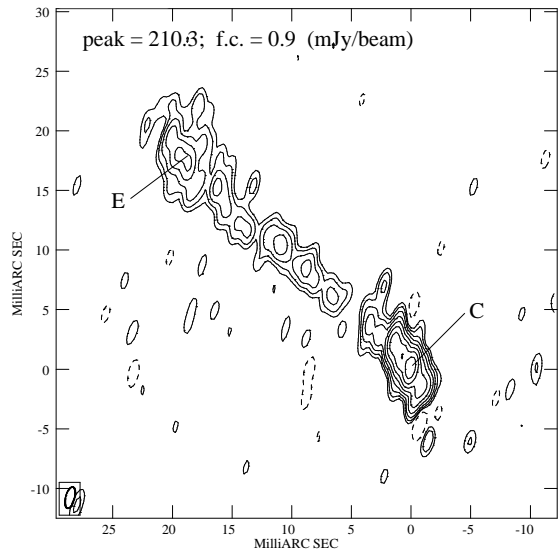}
\vspace{7.5cm}
\caption{{\bf 1413+349:} global VLBI image at 5 GHz. The
  restoring beam is 1.8$\times$0.7 mas$^{2}$ in p.a. $-14^{\circ}$,
  and is plotted in the bottom left corner. On the image we provide
  the peak flux density in mJy/beam, and the first contour (f.c.)
  intensity (mJy/beam), which corresponds to three times the
  off-source noise level. Contour levels increase by a factor of 2.}
\label{fig1413}
\end{center}
\end{figure}

\subsubsection{1413+349~~[~EF, ~$m$=..., ~z=...]} 

The radio source 1413+349 (alias OQ\,323) lacks an optical
identification. At 5 GHz the radio source displays a core-jet
structure. The counter-jet, visible in the 1.7-GHz
image (component West in Dallacasa et al. 1995), is 
completely resolved out at this frequency.
The morphology presented in Fig. \ref{fig1413}, 
is in good agreement with VLBA
observations at the same frequency performed by \citet{helmboldt07}.
The radio emission is dominated
by the brightest compact component, labelled C in Fig. \ref{fig1413},
that accounts for 607 mJy (i.e. 66\% of the total VLA flux density)
and likely harbours the source core.
The well-collimated jet, labelled S in Fig. \ref{fig1413}, and a
hint of the counter-jet emerge from the 
core component. 
The jet is about 54 mas in size and is resolved in several compact
sub-components. It shows a peak at the outer NE edge of the
structure, where a hotspot is likely present. The
flux-density ratio between the eastern and western components is
estimated from the 1.7-GHz data and is $R_{\rm S} \sim 6.2$, while
the arm-length ratio is $R_{\rm R} \sim 3$, indicating a
brighter-when-farther behaviour, as expected if the eastern and
western components are the jet and counter-jet, respectively.\\   
The {\it total} spectral index of the core region is rather flat with 
$\alpha \sim 0.3$, while the jet has a steeper spectrum 
with $\alpha \sim 0.7$ (Fig. \ref{spix}h).\\
Our 5-GHz global-VLBI observations could recover only 93\% of the total flux
density measured by the VLA at this frequency. The missing flux
density is likely due to the western component which could not be
accounted for by these observations.\\
No significant polarized emission has been detected by the VLA at 5 GHz
($p <$ 0.4\%).\\

\begin{figure*}
\begin{center}
\includegraphics{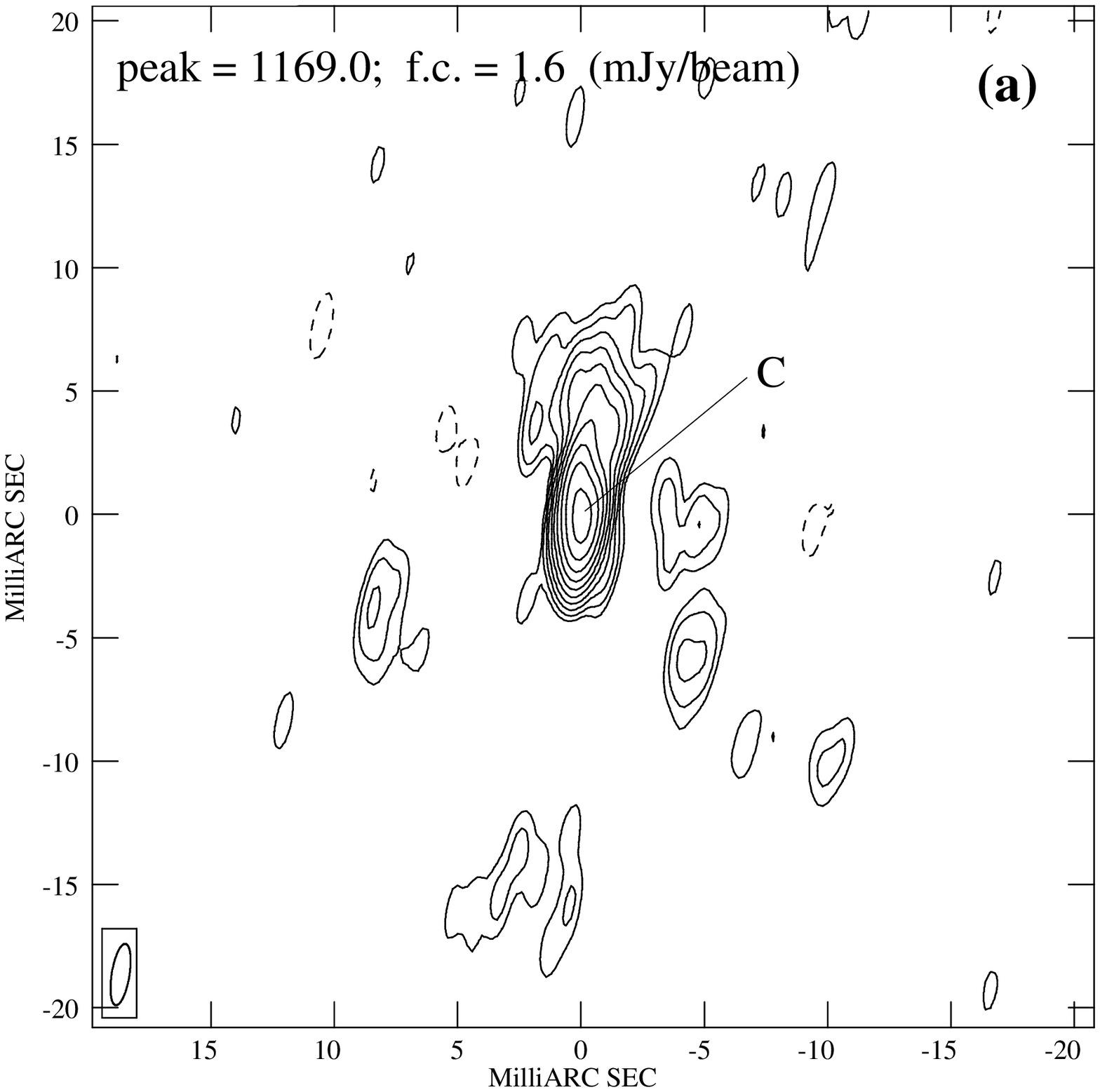}
\includegraphics{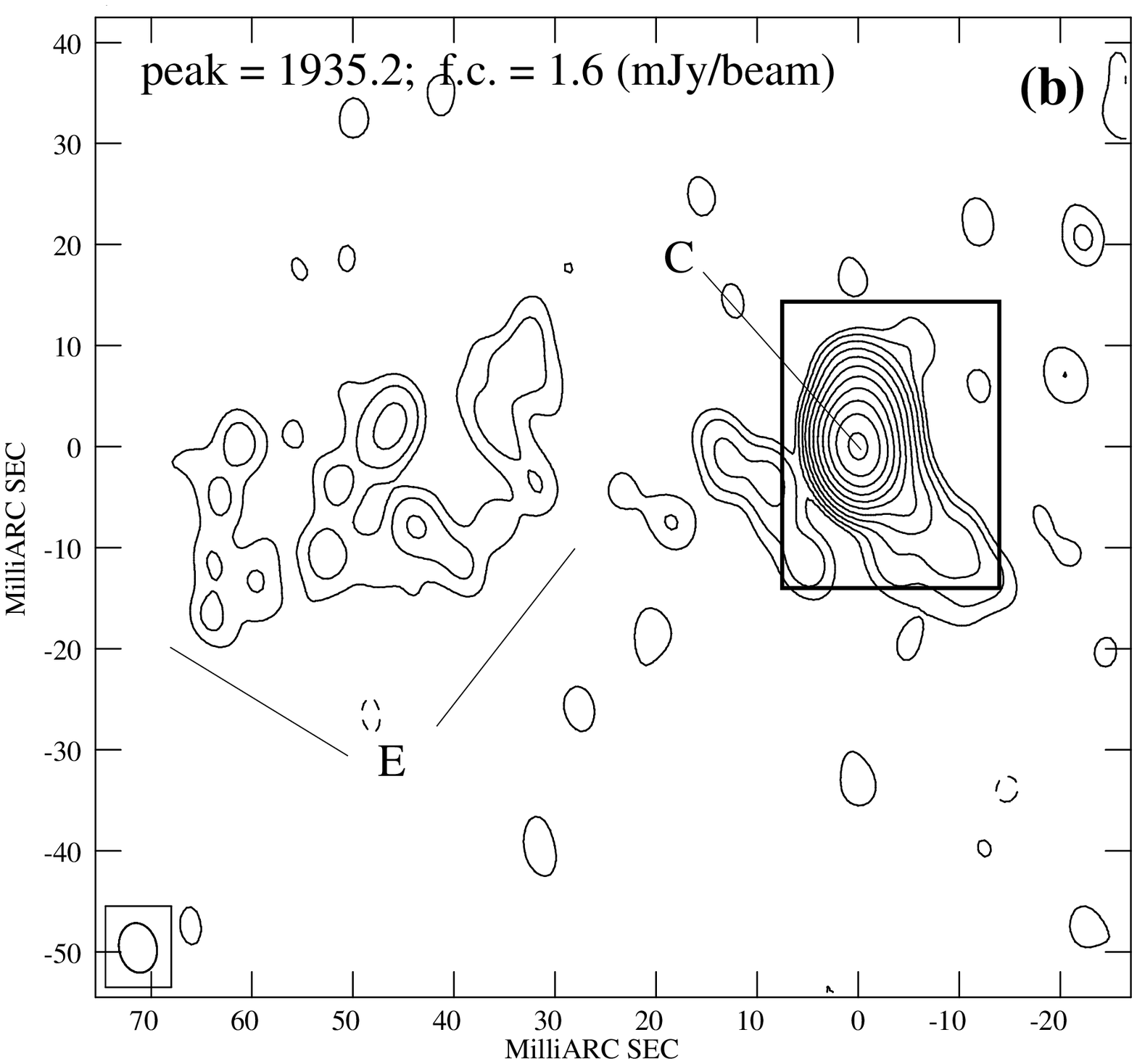}
\vspace{7.5cm}
\caption{{\bf 1600+335:} global VLBI image of the central component
  {\bf (a)} and of the total structure {\bf (b)}. 
In the first image the restoring beam is 2.5
  $\times$0.7 mas$^{2}$ in p.a. $-9^{\circ}$, and is plotted in the bottom
  left corner. The second image has been produced
with the short baselines only, and restored with
an elliptical beam of 5.0$\times$3.7 mas$^{2}$ in
p.a. 11$^{\circ}$. On each image we provide the peak flux density in
mJy/beam, and the first contour (f.c.) intensity (mJy/beam), which
corresponds to three times the off-source noise level. Contour levels
increase by a factor of 2. The box in panel b represents the area
  shown with higher resolution 
in panel {\bf (a)}.}  
\label{fig1600}
\end{center}
\end{figure*}

\subsubsection{1600+335~~[~G, ~$m_R$=23.0, ~z=...]} 

The radio source 1600+335 (alias 4C\,33.38) is optically identified
with a faint galaxy, but no spectroscopic redshift
is available so far  \citep{stickel96}. 
The redshift $z=1.1$ reported in \citet{snellen00} was
estimated on the basis of optical magnitudes, but no detail on
this approach was presented. For this reason we do not provide
any value for the redshift of this source.
The morphology of the radio emission is rather complex: the main
region, labelled C in Fig. \ref{fig1600}a, is resolved into a
sort of core-jet structure elongated in the NS 
direction, surrounded by 
blobs of the extended low-surface brightness emission visible at 1.7 GHz
and almost completely resolved out by these observations. 
Hints of component E are detected in the
low-resolution image only (Fig. \ref{fig1600}b).
The tapered image with the spectral
index information (Fig. \ref{spix}i) reveals an extension to the South
of the main region that bends 
towards the diffuse component located 40 mas to the East.
Since both the extended components detected in \citet{dd95} are severely
resolved in these 5-GHz observations, a reliable determination of the
spectral index could be done on the main component only, where it
turned out to be flat ($\alpha \sim 0.0$), supporting
the idea that it hosts the source core and the jet base, as suggested
by its morphology. \\
Our global VLBI image accounts for 92\% of the total flux density
measured by the VLA.\\
No significant polarized emission has been detected by the VLA at 5 GHz
($p <$ 0.5\%).\\

\begin{figure}
\begin{center}
\includegraphics{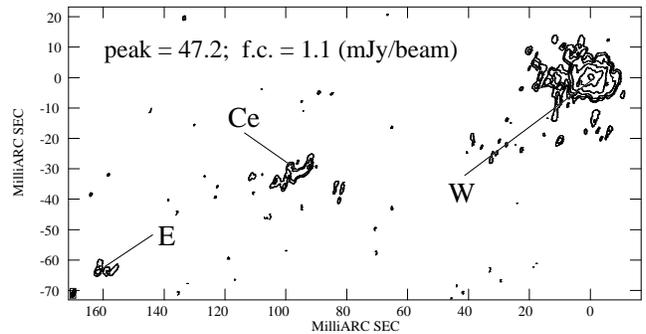}
\vspace{5cm}
\caption{{\bf 2342+821:} global VLBI image at 5 GHz. 
The restoring beam is 2.4$\times$ 0.8 mas in p.a. $-$19$^\circ$,
and is plotted in the bottom left corner. On the image we provide the
peak flux density in mJy/beam, and the first contour (f.c.) intensity
(mJy/beam), which corresponds to three times the off-source
noise level. Contour levels increase by a factor of 2.}
\label{2342}
\end{center}
\end{figure}

\subsubsection{2342+821~~[~Q, ~$m_R$=20.2, ~z=0.735]} 

The radio source 2342+821 is optically identified with a quasar
at redshift $z=0.735$ \citep{stickel96}.
In our 5-GHz global-VLBI 
image (Fig. \ref{2342}), 
the source shows an aligned triple structure of about 160 mas
(1.16 kpc) in size and position angle $\sim$110$^{\circ}$, 
in agreement with the morphology found at 1.7 GHz
\citep{dd95}. The radio
emission is dominated by the western component, W, that accounts for
1144 mJy that is 97\% of the total VLA flux density. These
observations could resolve the structure of components Ce and E. In
particular, component Ce is elongated in the same direction
of the whole source. 
The spectral index image (Fig. \ref{spix}j) does not reveal any region
with a flat/inverted spectrum, leaving the core identification an
open question. 
Component W has a spectral index $\alpha \sim$
0.8, while components Ce and E have steeper spectra.\\
Almost 98\% of the total flux density measured by the VLA is recovered
by our global VLBI observations.\\
No significant polarized emission has been detected by the VLA at 5 GHz
($p <$ 0.7\%).\\

\begin{figure*}
\begin{center}
\includegraphics{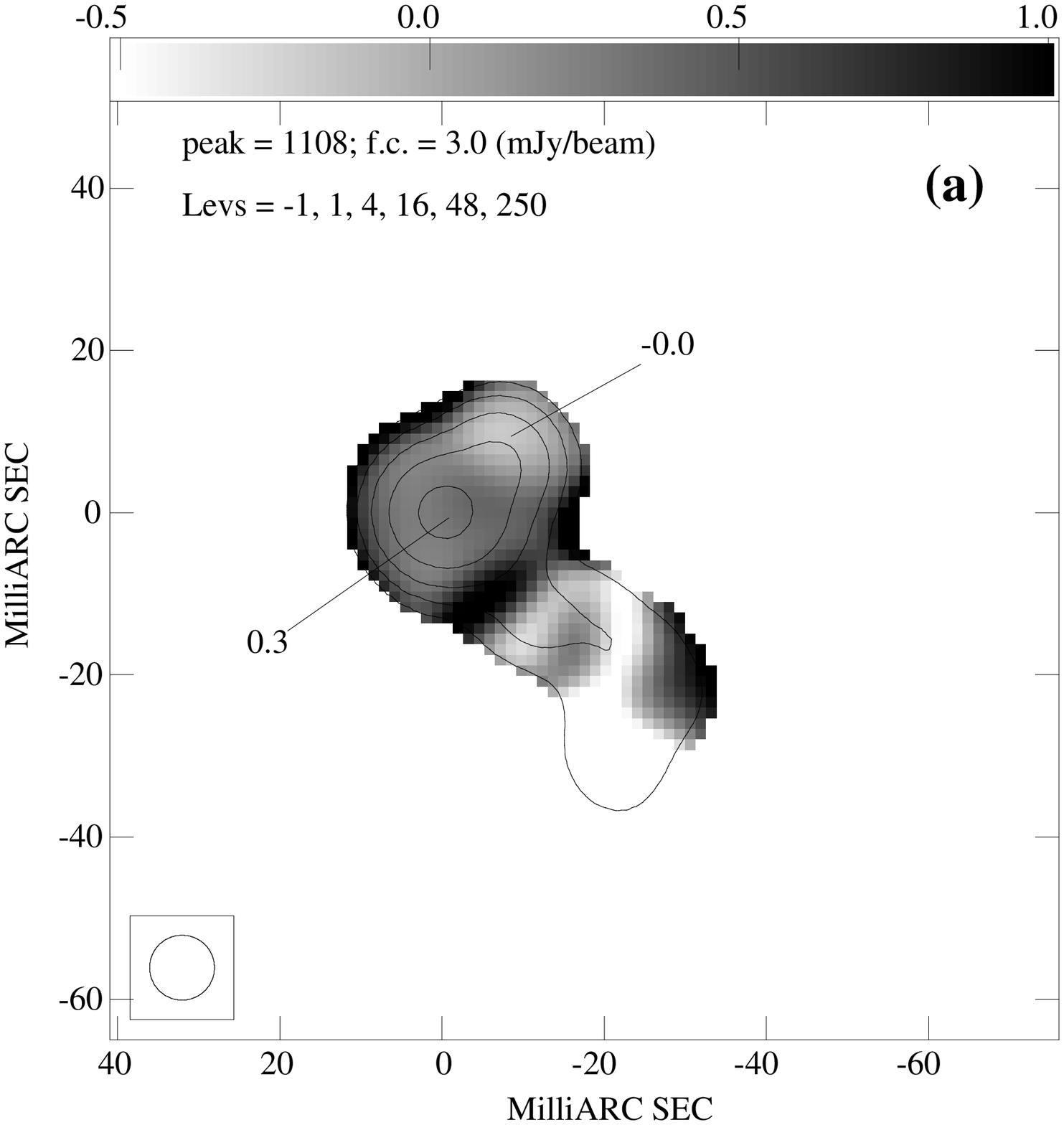}
\includegraphics{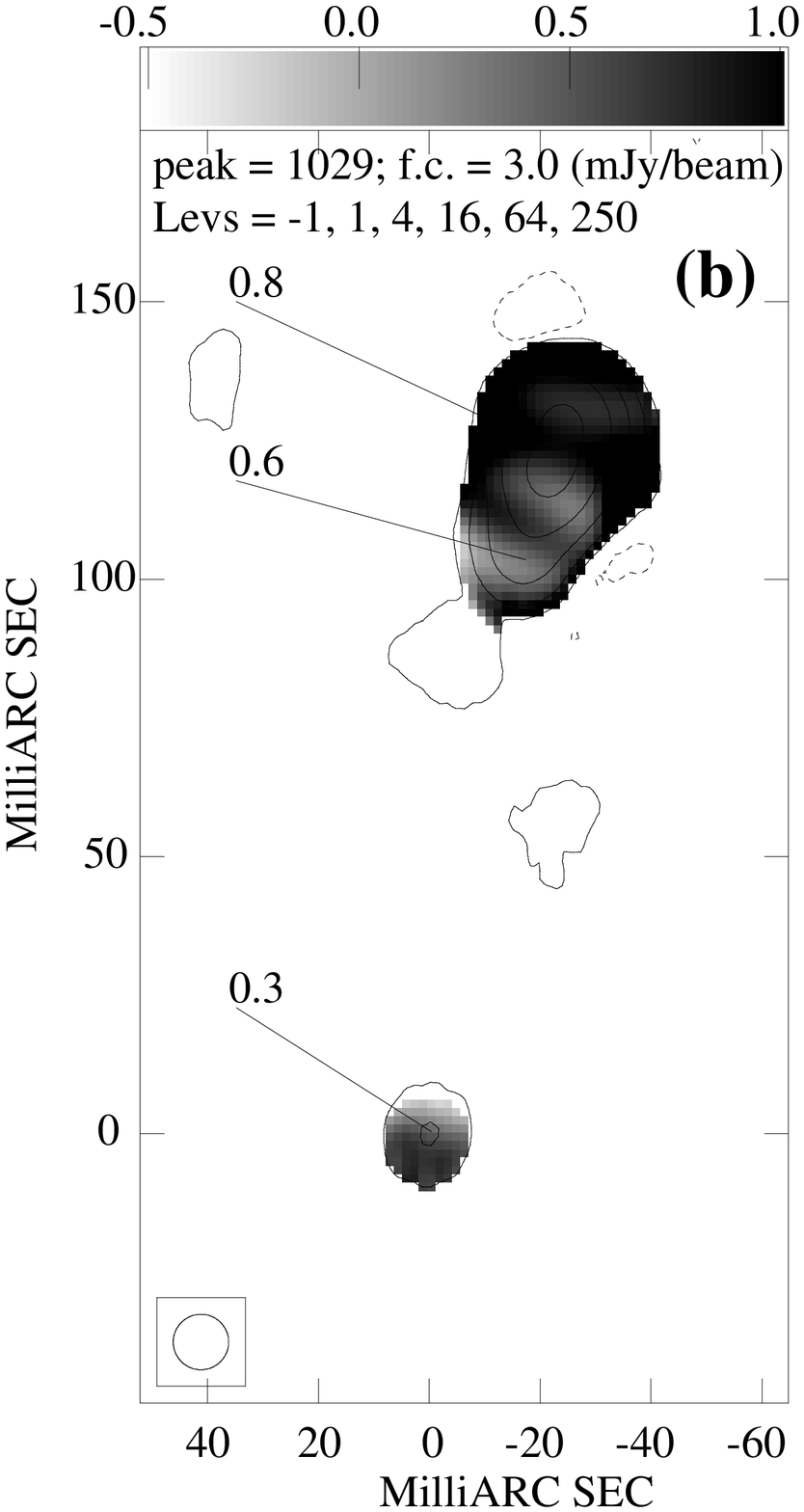}
\includegraphics{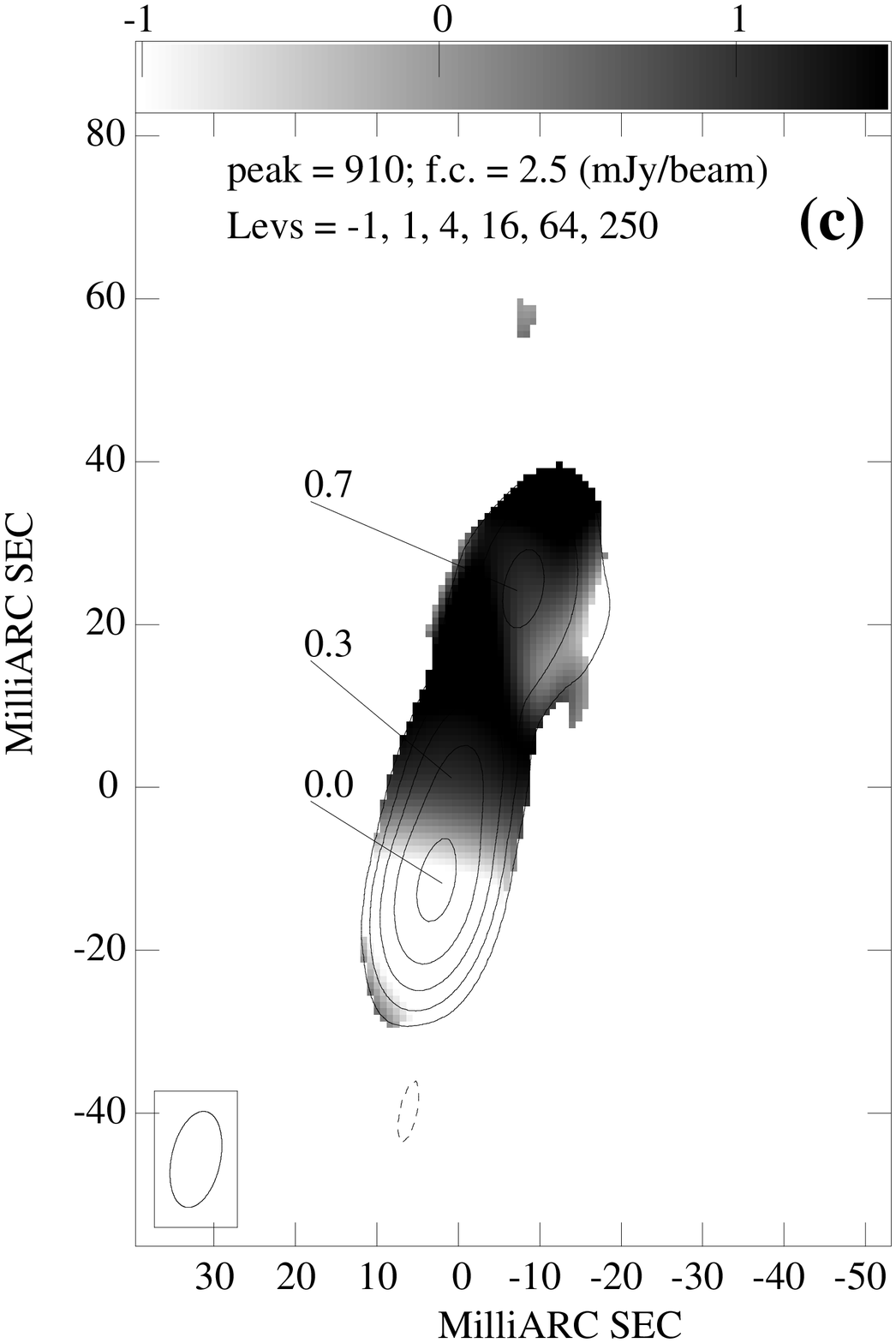}
\includegraphics{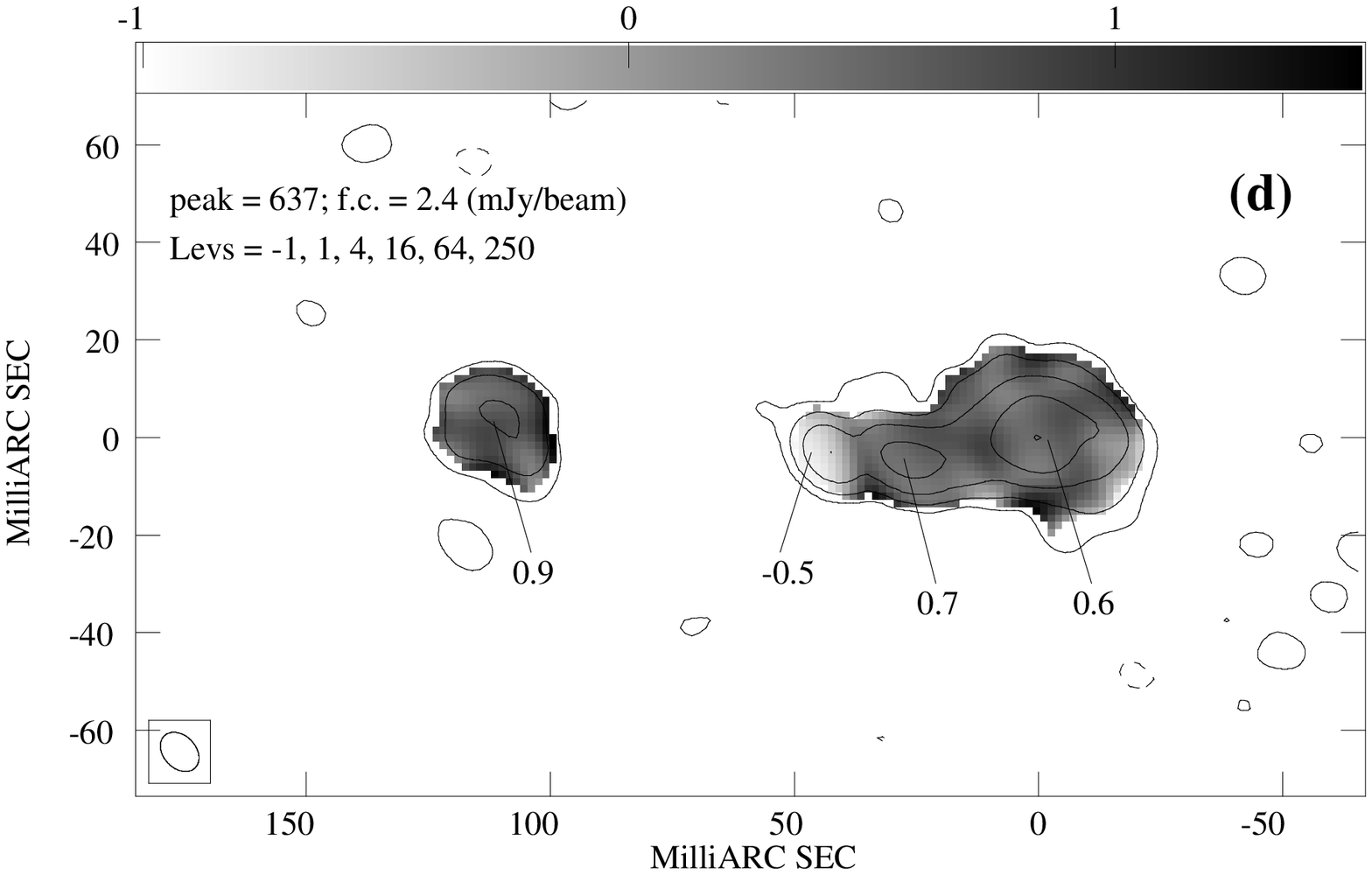}
\includegraphics{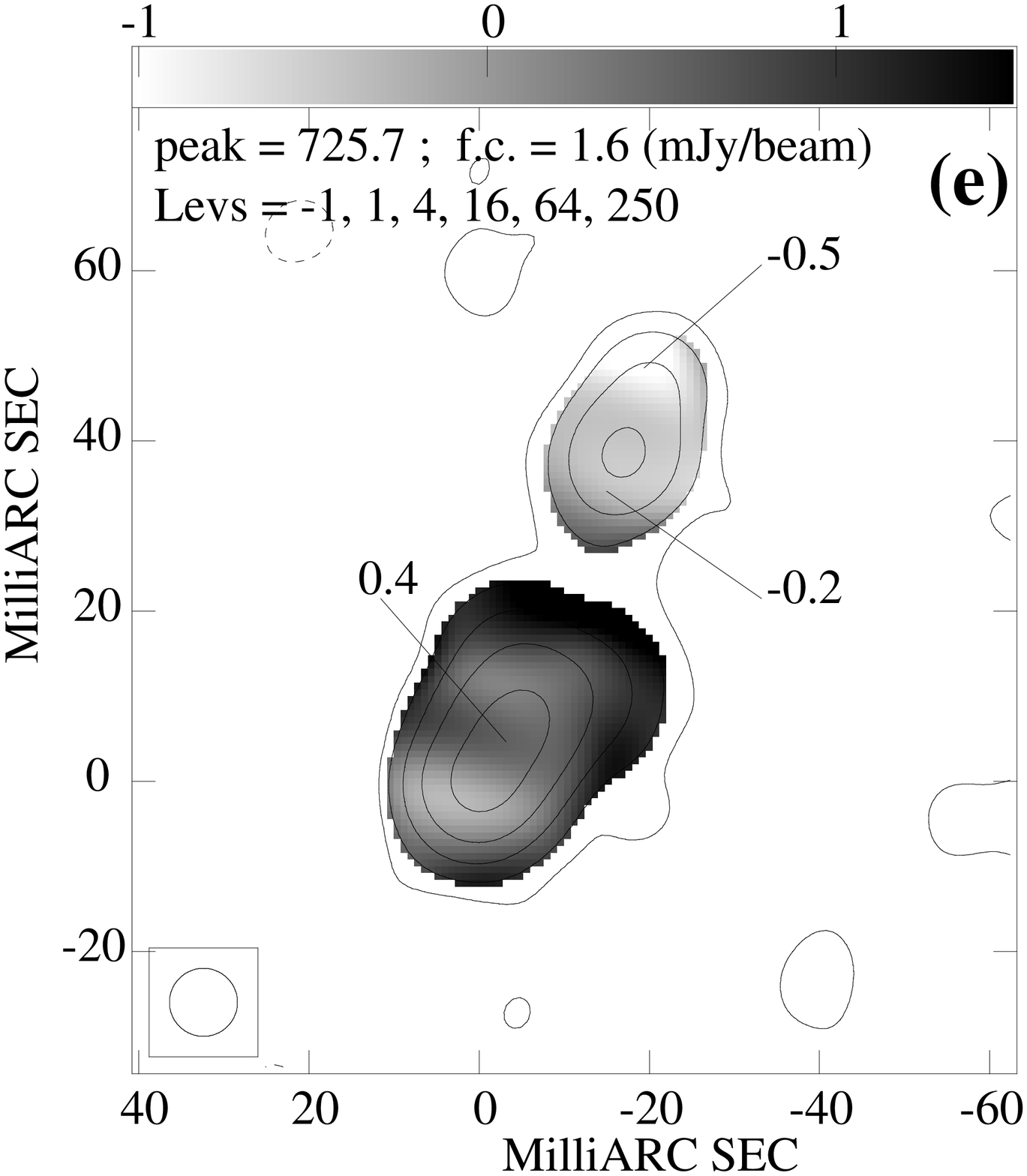}
\includegraphics{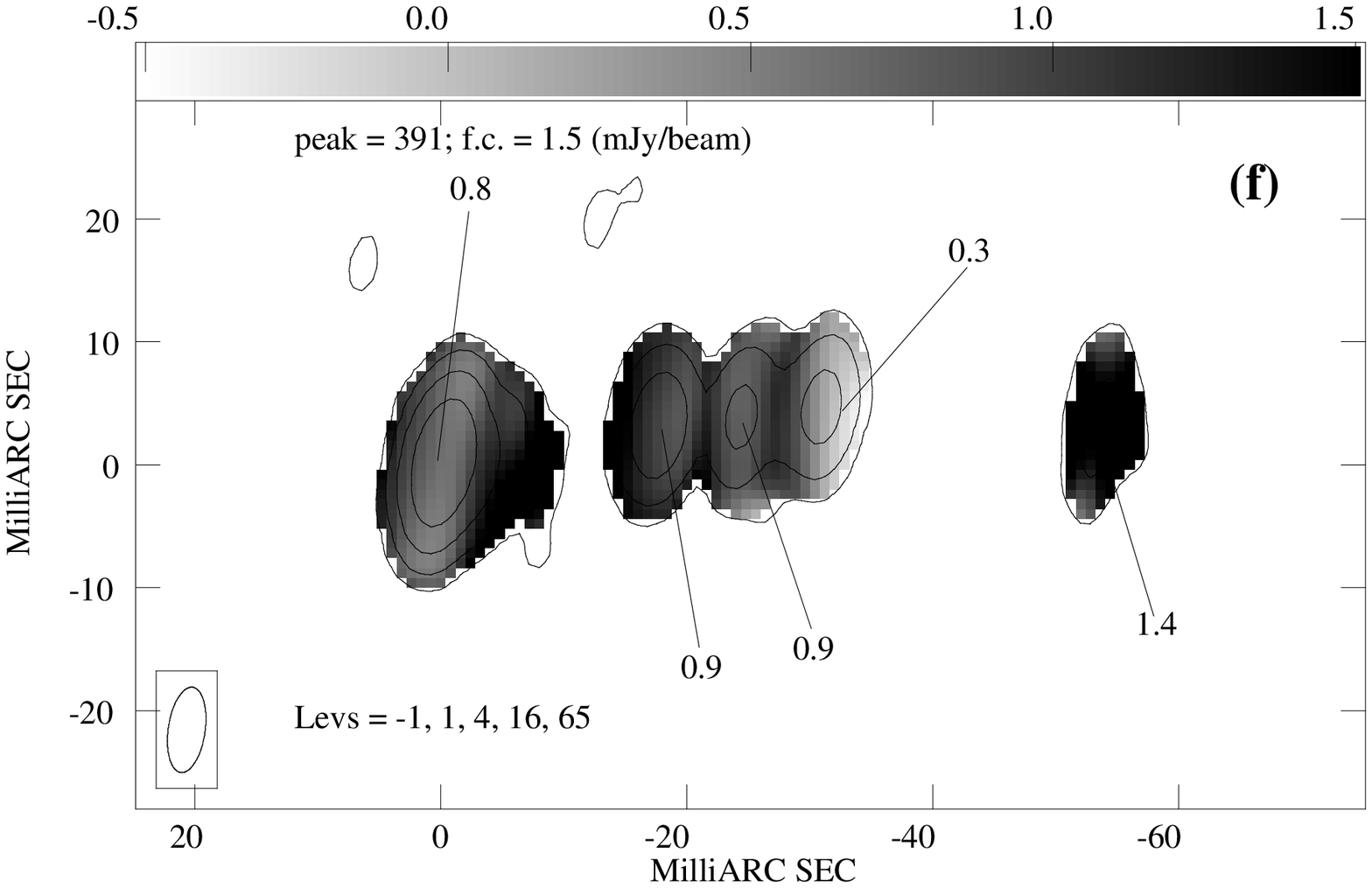}
\includegraphics{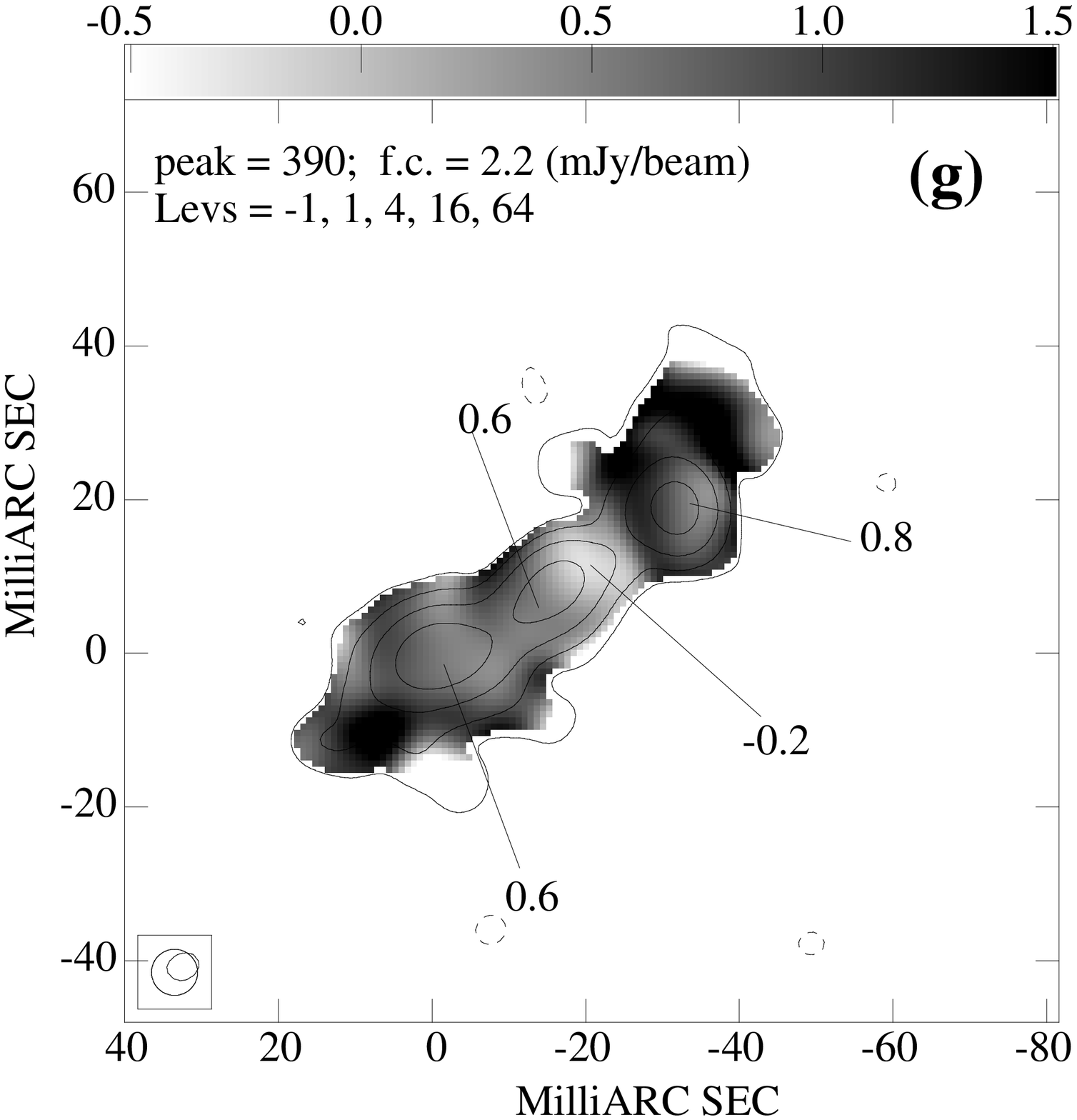}
\vspace{22cm}
\caption{Grey-scale spectral index images between 5.0 and 1.7 GHz of the sources
  studied in this paper, superimposed on the low-resolution 5-GHz
  contours convolved with the 1.7-GHz beam. {\bf (a): 0223+341; (b):
    0316+161; (c): 0319+121; (d): 0404+768; (e): 0428+205; (f): 1225+368; (g): 1358+624.}
  On each image we report the source name, the
  peak flux density in mJy/beam; the first contour (f.c.) intensity in
mJy/beam, the increment contour factor, and the restoring beam plotted
on the bottom left corner. The grey scale is shown by the wedge at the
top of each spectral-index image. }
\label{spix}
\end{center}
\end{figure*}

\addtocounter{figure}{-1}
\begin{figure*}
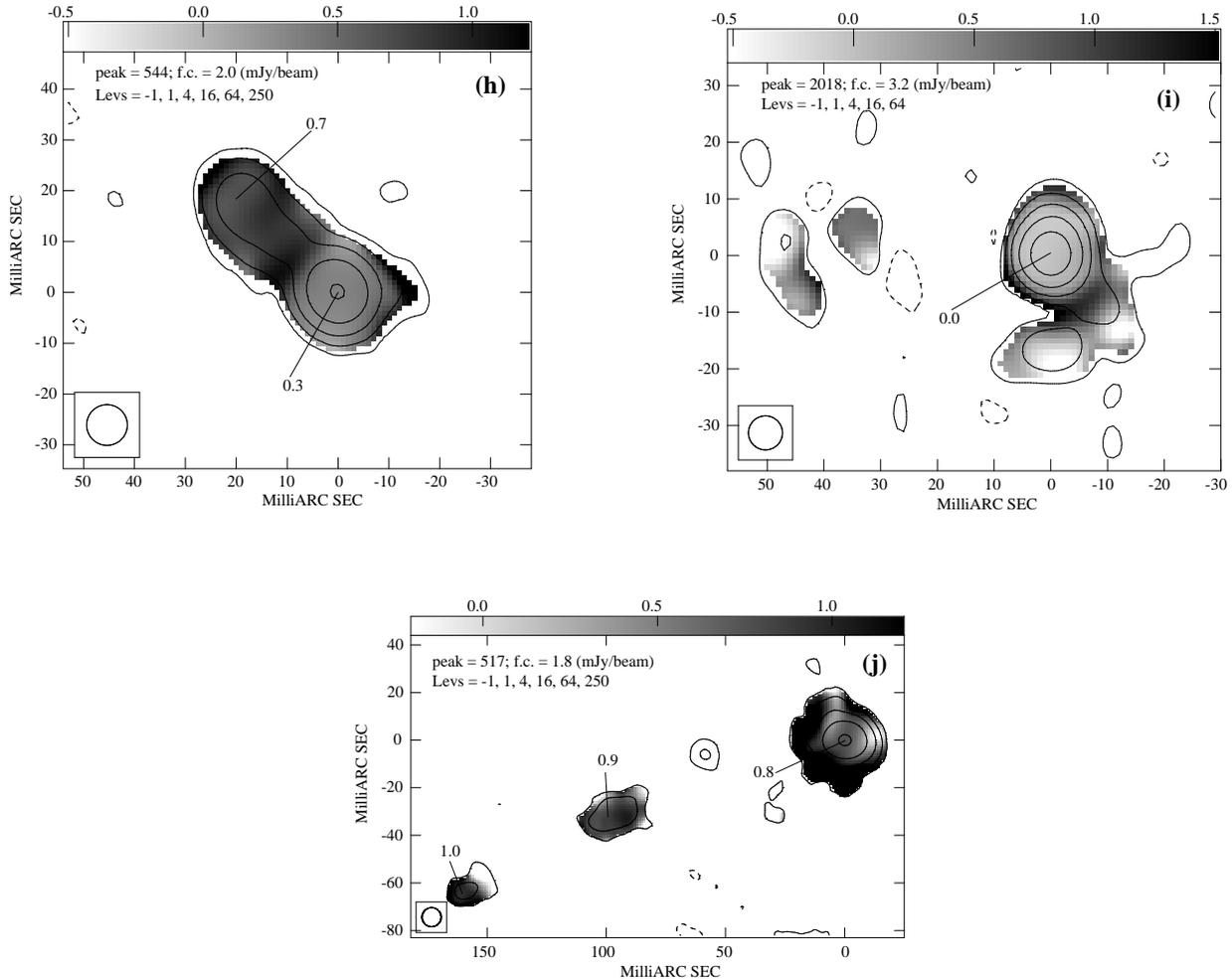

\begin{center}
\includegraphics{n1413_spixh.eps}
\includegraphics{n1600_spixi.eps}
\includegraphics{n2342_spixj.eps}
\vspace{14cm}
\caption{Continued. {\bf (h): 1413+349; (i): 1600+335; (j): 2342+821.}}
\end{center}
\end{figure*}

\section{Discussion}

\begin{table}
\caption{Morphology information.}
\begin{center}
\begin{tabular}{lcccc}
\hline
  & Q & G & EF & Tot \\
\hline
Two-sided  & 3 & 4 & 1 & 8\\
Core-jet   & 1 & 0 & 0 & 1\\
Complex    & 0 & 1 & 0 & 1\\
Core\tablenotemark{a}       & 4 & 2 & 1 & 7\\
Core-dominated\tablenotemark{a} & 1 & 0 & 1 & 2\\
Lobe-dominated & 1 & 3 & 0 & 4\\
Jet-dominated  & 1 & 1 & 0 & 2\\
Brighter-closer & 0 & 3 & 0 & 3\\
Brighter-farther & 1 & 1 & 1 & 3\\
\hline
\end{tabular}
\end{center}
\tablenotetext{a}{Sources with an unambiguous core identification.}
\label{morpho}
\end{table}

\subsection{Radio morphology}

The high spatial resolution images provided by our global VLBI observations,
complemented with the spectral index information, allow us to properly
describe the morphology of the CSS sources studied in this paper.\\
Among the 10 target sources, eight (3 quasars 0223+341, 
1225+368, and 2342+821; 4 galaxies 0316+161, 0404+768, 0428+205, and
1358+624; 1 empty field 1413+349) have a two-sided structure, one
source (the quasar 0319+121) has a one-sided core-jet morphology, while
the galaxy 1600+335 shows a complex structure. \\
In seven sources (0316+161, 0319+121, 0404+768, 0428+205, 1225+368,
1358+624, and 1413+349) the core region is unambiguously detected,
while in 0223+341 high-resolution observations at 327 MHz suggest
that the source core is hosted in the eastern component.
In general the radio emission is dominated either by lobes or by jets, while
the core, when detected, usually accounts for about 1\% of the
source flux density, with the exception of two objects (the quasar 0319+121,
and the empty field 1413+349)
where boosting effects may play a
major role. \\
In four sources (1 quasar 2342+821, and 3
galaxies 0316+161, 0404+768, and 0428+205) 
the radio emission comes mainly from the extended lobes, 
where the presence of a hotspot is 
indicated by a local flattening of the spectral index distribution.
Among the other sources, in 2 objects (1 quasar 1225+368, and 1 galaxy
1358+624) the jet is responsible for the majority of 
the source flux density. In the sources 1358+624 and 1413+349 
both the jet and a hint of the counter-jet
are visible. The edge-darkened structure at both 5 and 1.7 GHz of 1358+624
\citep{dd95} makes this source similar to
a scaled-down version of
FR\,I \citep{fr74} radio galaxy. As a comparison 
it is worth mentioning that
jets in CSS objects are usually edge-brightened.
The source 1600+335 shows a very complex structure dominated by a
compact component slightly resolved in the north direction
suggesting the presence of a jet, that is surrounded by extended
low-surface brightness emission almost completely resolved out at 5 GHz.\\  
A summary of the morphology information is reported in Table
  \ref{morpho}.\\ 
 
\subsection{Asymmetries}

\begin{table}
\caption{Luminosity and asymmetry parameters of the two-sided sources
  with an unambiguous core detection. Column 1: source name; Cols. 2,
  3: flux density ratio at 1.7 and 5 GHz, respectively; Col. 4:
  arm-length ratio; Cols. 5, 6: Source total luminosity and core
  luminosity, respectively.}
\begin{center}
\begin{tabular}{cccccc}
\hline
Source&$R_{\rm S, L}$&$R_{\rm S, C}$&$R_{\rm R}$&Log$L_{\rm tot}$&Log$L_{core}$\\
      &             &             &          & [W/Hz]        & [W/Hz]\\
(1)&(2)&(3)&(4)&(5)&(6)\\
\hline
&&&&&\\
0316+161&14.0&  - &0.6&28.51&25.87\\
0404+768& 9.5&7.0&0.6&28.08&26.43\\
0428+205& 6.0& -  &0.3&26.60&24.60\\
1225+368&12.8&20.3&1.4&28.46&27.23\\
1358+624& 2.2& 3.2&1.9&27.65&25.39\\
1413+349& 6.2&  -  &3.0&27.92&27.50\\
&&&&&\\
\hline
\end{tabular}
\label{asymmetry}
\end{center}
\end{table}

\begin{figure}
\begin{center}
\includegraphics{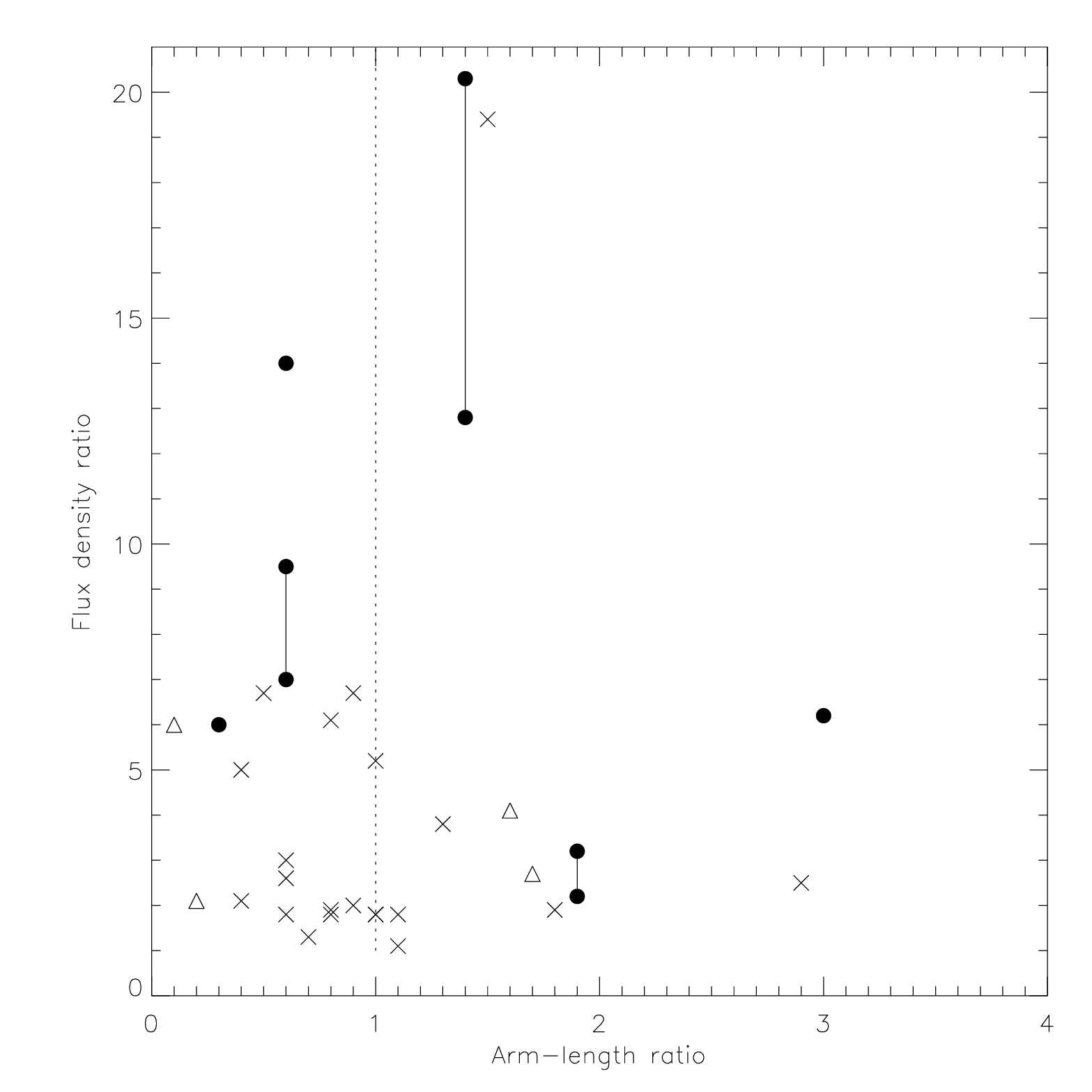}
\vspace{7cm}
\caption{Flux density ratio versus arm-length ratio for the six double
  sources with unambiguous core detection discussed in this paper 
({\it filled circles}), and
  for the CSS sources from
  \citet{rossetti06} ({\it crosses}), and \citet{mo04} ({\it
    triangles}). When the flux density ratio could be derived at both
  1.7 and 5 GHz, both values are plotted, and they are connected by a
  solid line. The vertical dotted 
line corresponds to an arm-length ratio of unity.}
\label{arm_figure}
\end{center}
\end{figure}

\begin{figure}
\begin{center}
\includegraphics{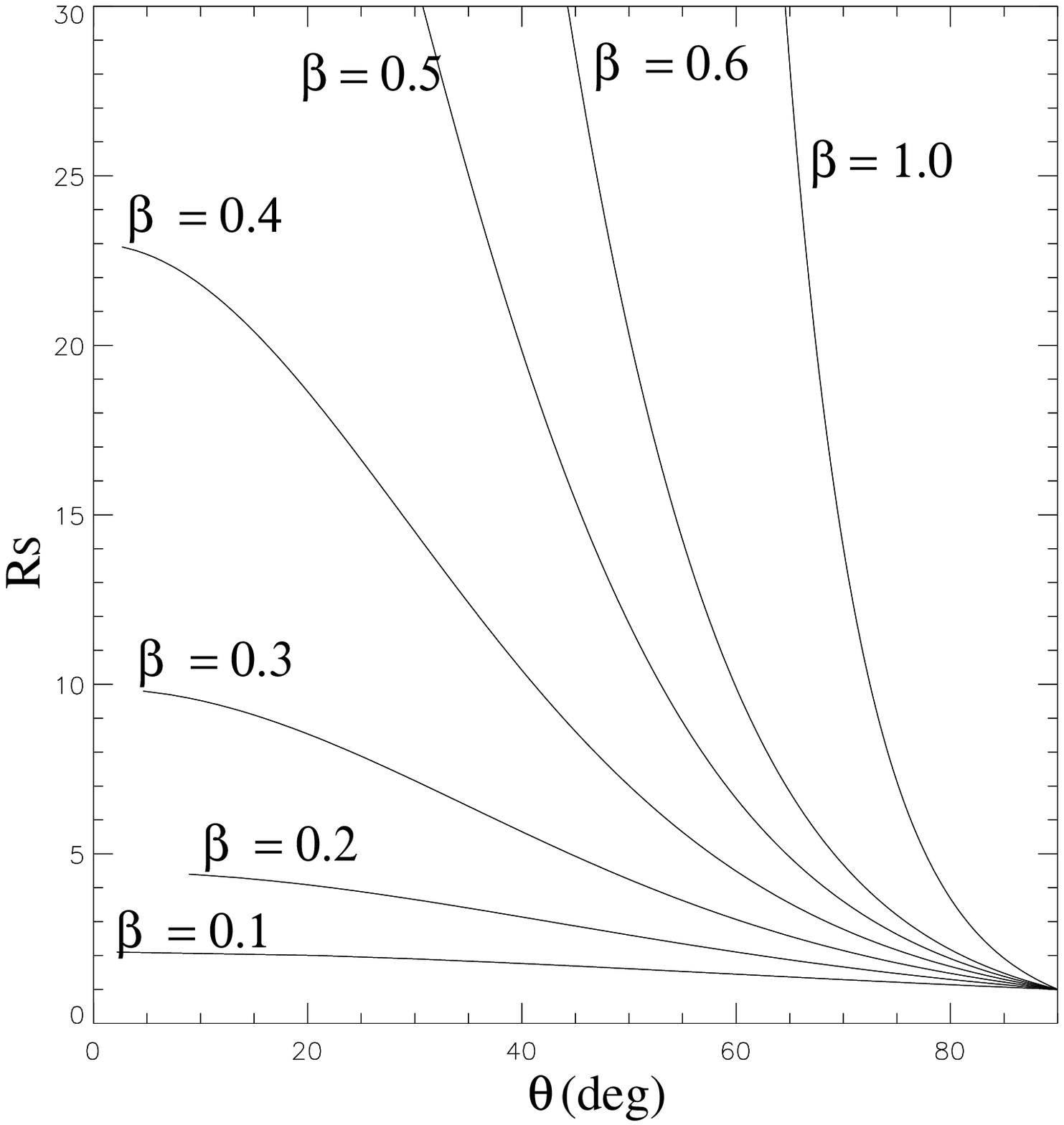}
\vspace{7cm}
\caption{Flux density ratio versus the orientation of the source axis,
as a function of the advance speed (see Eq. \ref{luminosity_ratio}).}
\label{r_theta}
\end{center}
\end{figure}

The radio structure of the six two-sided sources with a secure core
detection is not perfectly symmetric,
and the majority of the radio emission arises from one of the
jets/lobes. 
The flux density ratio $R_{\rm S}$ 
between the two sides
of the source ranges from 2.2 in 1358+624 to 20.3 in the most
asymmetric source 1225+368. In three sources, 1225+368,
1358+624 and 1413+349, 
we found a brighter-when-farther behaviour in which the
brightest lobe/jet is the farthest from the core, as expected from
projection and relativistic effects. 
In the case of the three galaxies 
0316+161, 0404+768 and 0428+205
the core region is closer to the
brighter lobe, which is the opposite to what is expected in the above
assumptions. In Table \ref{asymmetry}, the arm-length ratio and
  the flux density ratio are reported for the six two-sided sources
  with an unambiguous core detection.  \\
Interestingly, among the asymmetric radio sources presented here, HI
absorption has been found in all the four radio galaxies: 
0316+161, 0404+768, 0428+205, 
and 1358+624 \citep{salter10,vermeulen03}. 
However, the lack of high spatial
resolution observations does not allow us to locate the position of
the HI and thus no conclusive results on a possible jet-medium
interaction can be drawn. No observations searching
for HI absorption in 1225+368 and 1413+349 are available.\\
In Fig. \ref{arm_figure} we plot the flux-density ratio versus the arm-length
ratio for the two-sided CSS sources from this paper and from the CSS
sample selected by \citet{fanti01} and studied in details by
\citet{mo04} and \citet{rossetti06}. A $R_{\rm R} < 1$ indicates a
brighter-when-closer behaviour. Among the 31 two-sided CSS sources
considered, we found that 16 objects ($\sim$52\%) have a
brighter-when-closer behaviour. For these sources, 
the strong asymmetries in flux density and arm-length may be
caused by the interaction between one jet and the inhomogeneous
ambient medium as pointed out by several studies of asymmetric CSS/GPS sources 
\citep[e.g.][]{morganti04,labiano06}. 
The interaction between the advancing jet and a clumpy medium may
  enhance the luminosity due to high radiative losses which become
  predominant with respect to the adiabatic ones. The enhancement of
  the flux density may be at the 
origin of the high number counts of CSS/GPS objects in flux density limited
samples, which are larger than what is expected from the number counts
of classical radio galaxies.
Furthermore, although the jet-medium interaction 
may not frustrate the source expansion for its
whole lifetime, it may severely slow down its growth.  
The high fraction of CSS sources with a
brighter-when-closer behaviour suggests that jet-cloud interactions
are not so unlikely and it may cause an
underestimate of the source age. \\
Among the sources with the
brighter-when-farther behaviour, we investigate the advance speed and
the jet orientation assuming that both jets are thrust by the same
jet power and are expanding within a homogeneous medium. 
In this context, we assume a simple beaming model in which 
the asymmetries are produced by
boosting effects. The flux density ratio is:\\

\begin{equation}
R_{\rm S} = \left( \frac{1+ \beta {\rm cos} \theta}{1- \beta {\rm cos} \theta} \right)^{3 + \alpha}
\label{luminosity_ratio}
\end{equation}

\noindent where $\beta$ is the source advance speed and $\theta$ is
the jet orientation angle to our line of sight. We note that the
comparison of flux densities may not be appropriate due to
relativistic time dilation that causes different evolution between the
approaching and receding structures. However, this effect should not be
relevant in these sources because the lobes/hotspots are oriented at
larger angles than jets in blazars, and the evolution should be
smothered, as supported by the lack of flux density variability in
these sources.\\ 
By means of Eq. \ref{luminosity_ratio} we can derive the
possible combinations of $\beta$ and $\theta$ which reproduce 
the observed flux density asymmetries. 
Fig. \ref{r_theta} shows the flux density
ratio versus the jet orientation as a function of the advance
speed. In this case we conservatively assumed $\alpha = 0.7$. It is clear that a
source oriented at large angles to our line of sight, 
$\theta \geq 80^{\circ}$, 
cannot produce large asymmetries even in the case the jet is
advancing with $\beta = 1$. On the other hand, large asymmetries,
i.e. $R_{\rm S} > 10$, can be produced only assuming $\beta > 0.3$. \\
Among the sources studied in this paper, the object with the
highest flux density ratio is 1225+368 with $R_{\rm S}
=20.3$ and 12.8 at 5 and 1.7 GHz, respectively. 
Such high values can be explained assuming
an advancing velocity of $\beta \gtrsim 0.4$ and $\theta \gtrsim
30^{\circ}$. These are reasonable values in the case of 1225+368,
which is optically associated with a quasar.\\
The other source showing a high flux density ratio ($R_{\rm S} \sim 19$)
is the source B3\,0039+398 (alias 4C\,39.03)
from \citet{rossetti06}. Although no information on the optical
counterpart is available, the high flux density ratio suggests that
this source is likely associated with a quasar rather than a galaxy. 
According to the unified models, 
radio galaxies should be oriented
at $\theta > 45^{\circ}$, with the average value of $\theta =$70$^{\circ}$
\citep{barthel89}. In a galaxy, such a strong asymmetry can be obtained
only if the radio galaxy is expanding with $\beta > 0.5$. 
However, kinematic studies of young radio
galaxies indicate that the mean expansion velocity is
$\sim$0.1$c$ \citep{polatidis03,giroletti09}, implying that high
asymmetries in galaxies are unlikely, because other {\it ad hoc} 
ingredients should be
invoked in addition to projection and relativistic effects.\\

\subsection{Physical parameters}

\begin{figure}
\begin{center}
\includegraphics{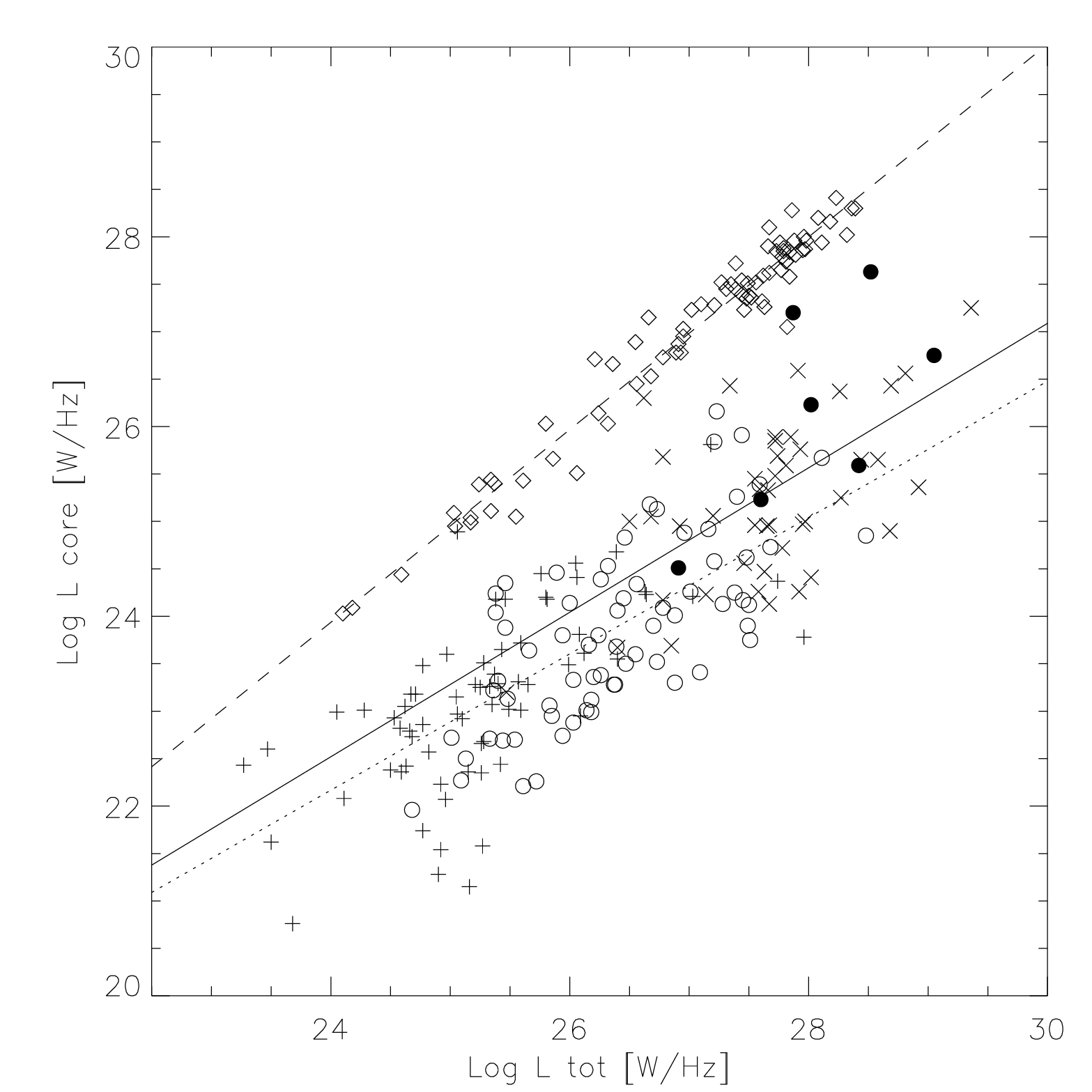}
\vspace{7cm}
\caption{Core luminosity at 5 GHz versus the total source
  luminosity at 408 MHz for FRI ({\it plus sign}) and FRII ({\it
    empty circles}) radio sources \citep{zirbel95}, 
CSS/GPS sources ({\it crosses})
\citep{fanti00,fanti90,stanghellini98,peck00,dd00}, blazars ({\it
  diamonds}) \citep{abdo09}, and the CSS sources studied in this paper
({\it filled circles}). The solid, dotted and dashed lines represent
the best linear fit for the CSS/GPS objects, FRI/FRII
objects, and blazars, respectively (see Section 4.3)}
\label{lum-lum}
\end{center}
\end{figure}

For the seven sources with a clear detection of the core component we
investigate the contribution of the core luminosity to the source
total luminosity and we compare the values with what was found for other
classes of objects. In Fig. \ref{lum-lum} we plot the luminosity of
the core at 5 GHz versus the source total luminosity at 408 MHz for a
sample of FRI and FRII from \citet{zirbel95}, and for CSS/GPS objects with a
secure detection of the core from the samples by 
\citet{fanti01,fanti90,peck00,stanghellini98,dd00}. 
For all the sources
lacking observations at 408 MHz, the flux density at such frequency
was extrapolated from the optically thin part of the
spectrum. The same approach was used to derive the flux density at 408
MHz for those CSS/GPS objects whose spectrum was already self-absorbed at this
frequency. In the case the flux density of the core at 5 GHz was not
available, the 5-GHz flux density was extrapolated assuming
a flat spectrum ($\alpha=0$). \\
As a comparison, we
included also the blazars from
the 3-month {\it Fermi}-LAT bright $\gamma$-ray source list
\citep[LBAS,][]{abdo09}. 
We note that the total flux density for the blazar population should be
considered a lower limit since it refers to observations at 0.8 or 1.4
GHz \citep{giroletti10} where the contribution of extended structures
is less effective than at 408 MHz. \\ 
From Fig. \ref{lum-lum} we found that CSS/GPS sources are the
high-luminosity tail of the FRII population. Among seven sources of our
sample with the core detected, five share the same region as FRII and
CSS/GPS, while two objects, 0319+121 and 1413+349 
fall in the area occupied by the
blazars. These outliers are objects whose radio emission is dominated
by the core/jets,  
and this suggests that boosting effects may play a dominant role in their
radio emission.\\
To derive a relation between the core and the total luminosity, we
performed a linear regression fit of Log $L_{\rm core}$ versus Log
$L_{\rm tot}$:\\

\begin{displaymath}
{\rm Log} L_{\rm core} = a + b \times {\rm Log} L_{\rm tot} 
\end{displaymath}

\noindent considering CSS/GPS
objects. As a comparison, we performed two additional linear regression fits
considering FRI/FRII and blazars, separately. 
Fit parameters obtained
minimizing the chi-square error statistic are reported in
Table \ref{fit}. 

\begin{table}
\caption{Parameters of the linear regression fit of log $L_{\rm core}$
versus log $L_{\rm tot}$.}
\begin{center}
\begin{tabular}{ccc}
\hline
Sources&a&b\\
\hline
&&\\
CSS/GPS      & 4.24$\pm$4.06&0.76$\pm$0.15\\
FRI/FRII     & 4.92$\pm$1.49&0.72$\pm$0.06\\
LBAS Blazars &-0.43$\pm$0.64&1.01$\pm$0.02\\
&&\\
\hline
\end{tabular}
\end{center}
\label{fit}
\end{table}

\noindent The slope obtained for CSS/GPS sources is a little
  steeper than that for FRI/FRII, although within the uncertainties,
 suggesting a lower contribution from the extended structures.
Another possibility is that the CSS/GPS sources have a slightly larger
core emission. This may be due either to an
intrinsically brighter core, or 
to the additional contribution from the jet base that cannot be
disentangled from the core emission.
The slope obtained for FRI/FRII sources is in agreement
with what was found by \citet{gg01} for
a different sample of FRI/FRII objects.  
In the case of the blazar population from the LBAS, the slope is close to
1, indicating that the
contribution from extended structures is negligible and the core
dominates the source emission due to Doppler beaming effects, as it
was already pointed out by \citet{giroletti10}. \\
On the basis of the observed parameters, and assuming minimum energy
conditions (particle energy is equal to the magnetic energy), we
derive the physical properties for each
component of the radio sources. 
We computed the 
equipartition magnetic
field $H_{eq}$, the minimum energy density $u_{min}$, and the minimum
internal pressure $p_{min}$ using standard formulae \citep{pacho70}.
We assumed that the source components have an ellipsoidal volume
with a filling factor of unity, and an equal distribution between
proton and electron energies. 
An average optically-thin spectral index
of 0.7 has been adopted. We found that the
minimum energy density $u_{\rm min}$ is between 10$^{-4}$ and
10$^{-6}$ erg cm$^{-3}$, the minimum internal pressure $p_{\rm
  min}$ ranges between 10$^{-4}$ and 10$^{-7}$ dyne cm$^{-2}$, while the
equipartition magnetic field accounts for a few mG up to about 120 mG
(Table \ref{component}).
The latter value has been found in the compact core of the core-jet radio 
source 0319+121, which is also at very high redshift. The higher
values are found in compact components and are similar to those found
in the hotspots and core region of other CSS sources
\citep{fanti90,dd02,mo04}, whereas in extended components, like lobes and
jets, we found lower values.\\
 
\subsection{Polarization properties}

Several studies of the polarization properties of CSS sources
\citep[e.g.][]{cotton03,cf04} have shown that sources smaller than a
few kpc have very
low values of polarized emission at frequencies below 8.4 GHz. 
This result has been interpreted by
assuming that compact sources are highly depolarized by the dense
interstellar medium that enshrouds the radio emission and acts as a
Faraday screen. Support to this idea is the detection of some level of
polarized emission as we consider higher frequencies, 
while the sources are
completely unpolarized at the lower frequencies.\\
From our VLA observations, we found that 9 out of the 10 sources
studied here are unpolarized at 5 GHz, in good agreement with the expectation. 
The radio source 1358+624, unpolarized in our 5 GHz observations,
turned out to be polarized at 15 GHz
\citep{aller85}, suggesting that strong depolarization is taking place
in this source. \\
Only the quasar 0319+121 
turned out to be significantly polarized, with a fractional
polarization of about 5.5\%. 
Since polarimetric observations are available at one
frequency only, we cannot determine the rotation measure (RM). However, if
we assume a low RM, we obtain that the magnetic field is parallel to
the VLBI jet axis, as it is observed in quasars. The presence of
significant percentage of polarized emission, together with the
core-jet structure, suggests that this source may be oriented at a
small angle with respect to the line of sight and its linear
size may be foreshortened by projection effects.\\

\section{Summary}

We presented the results from global-VLBI observations at 5 GHz
of a sample of
10 out of 16 CSS sources from the sample presented by
\citet{dd95}. The majority of the targets have a two-sided structure
and the radio emission is dominated by lobes, jets, and
hotspots. One sources, 0319+121, has a core-jet morphology and
the radio emission comes mainly from the core component. This object
is the only one with significant polarized emission. High polarization
level is uncommon in CSS sources, where the interstellar
medium enshrouding the radio source acts as a Faraday screen causing
severe depolarization of the radio emission.\\
The determination of the nature of the source components has
been provided by the availability of two-frequency observations which
allowed the study of the spectral index distribution across the
whole source. The core has been unambiguously identified in 7
sources. Among them, the six sources with a two-sided structure have large
flux density and/or arm-length asymmetries. In three cases the
brightest lobe is the farthest from the core, in agreement with
what is expected in case of projection effects. 
The evidence that the ambient medium
may play a decisive role in producing asymmetric objects is supported
by the galaxies 0316+161, 0404+768 and 0428+205, where the brightest
lobe is also the closest to the core. The interaction with a dense
medium may slow down the jet growth, lowering adiabatic losses and
enhancing radiative losses. As a consequence the synchrotron emission
from the interacting jet is enhanced.
If the jet-environment
interaction is not a sporadic phenomenon in the lifetime of CSS radio
sources, it may provide a possible explanation of the high fraction of
CSS objects in flux density limited sample.\\ 
The influence of the ambient medium on the source expansion has
  been proved on a few individual objects only and no statistical
  study has been performed so far. Multi-frequency VLBA observations
  of a sample of the most compact CSS/GPS objects (with linear size of
  a few parsecs) have been already performed in order to investigate
  the role played by environment during the very first stage of the
  source growth. The results will be used 
  to draw a complete and reliable picture of the
  radio source evolution. 

\section*{Acknowledgments}

We thank the anonymous referee for reading the manuscript carefully
and making valuable suggestions. 
We thank the European VLBI Network and the United States VLBI Network
for carrying out the observations. The WSRT is operated by the
Netherlands Foundation for Research in Astronomy with the financial
support of the Netherlands Organization for Scientific Research (NWO). 
The National Radio Astronomy Observatory is operated by Associated
Universities, Inc., under contract with the National Science
Foundation.  This research has made use of the NASA/IPAC Extragalactic
Database (NED), which is operated by the Jet Propulsion Laboratory,
California Institute of Technology, under contract with the National
Aeronautics and Space Administration. DD acknowledges the Commission
of the European Union for the award of a Fellowship.
Funding for SDSS-III has been provided by the Alfred P. Sloan Foundation, the Participating Institutions, the National Science Foundation, and the U.S. Department of Energy Office of Science. The SDSS-III web site is http://www.sdss3.org/.

SDSS-III is managed by the Astrophysical Research Consortium for the Participating Institutions of the SDSS-III Collaboration including the University of Arizona, the Brazilian Participation Group, Brookhaven National Laboratory, University of Cambridge, Carnegie Mellon University, University of Florida, the French Participation Group, the German Participation Group, Harvard University, the Instituto de Astrofisica de Canarias, the Michigan State/Notre Dame/JINA Participation Group, Johns Hopkins University, Lawrence Berkeley National Laboratory, Max Planck Institute for Astrophysics, Max Planck Institute for Extraterrestrial Physics, New Mexico State University, New York University, Ohio State University, Pennsylvania State University, University of Portsmouth, Princeton University, the Spanish Participation Group, University of Tokyo, University of Utah, Vanderbilt University, University of Virginia, University of Washington, and Yale University.


\begin{thebibliography}{}

\bibitem[Ahn et al.(2012)]{ahn12}
Ahn, C.P., et al. 2012, ApJS, 203, 21

\bibitem[Abdo et al.(2009)]{abdo09}
Abdo, A.A., Ackermann, M., Ajello, M., et al. 2009, ApJS, 183, 46 

\bibitem[Aller et al.(1985)]{aller85} 
Aller, H.D., Aller, M.F., Latimer G.E., Hodge P.E. 1985, ApJS, 59, 513.  

\bibitem[Barthel(1989)]{barthel89}
Barthel, P.D. 1989, ApJ, 336, 606

\bibitem[Clark(1973)]{clark73} 
Clark, B.G. 1973, Proc. IEEE, 61, 1242

\bibitem[Cohen et al.(1975)]{cohen75} 
Cohen, M.H., et al. 1975, ApJ, 201, 249

\bibitem[Conway(1999)]{conway99}
Conway, J.E. 1999, NewAR, 43, 509

\bibitem[Cotton et al.(2003)]{cotton03}
Cotton, W.D., et al. 2003, PASA, 20, 12

\bibitem[Dallacasa et al.(1995)]{dd95} 
Dallacasa, D., Fanti, C., Fanti, R., Schilizzi, R.T.,
Spencer, R.E., 1995, A\&A, 295, 27

\bibitem[Dallacasa et al.(2000)]{dd00}
Dallacasa, D., Stanghellini, C., Centonza, M., Fanti, R. 2000, A\&A, 363, 887

\bibitem[Dallacasa et al.(2002)]{dd02}
Dallacasa, D., Tinti, S., Fanti, C., Fanti, R., Gregorini, L.,
Stanghellini, C., Vigotti, M. 2002, A\&A, 389, 115

\bibitem[Fanaroff \& Riley(1974)]{fr74}
Fanaroff, B.L., Riley, J.M. 1974, MNRAS, 167, 31

\bibitem[Fanti et al.(1990)]{fanti90}
Fanti, R., Fanti, C., Schilizzi, R.T., Spencer, R.E., Nan, R., Parma, P., van Breugel, W.J.M., Venturi, T. 1990, A\&A, 231, 333

\bibitem[Fanti et al.(1995)]{fanti95}
Fanti, C., Fanti, R., Dallacasa, D., Schilizzi, R.T., Spencer, R.E.,
Stanghellini, C. 1995, A\&A, 302, 31 

\bibitem[Fanti et al.(2000)]{fanti00}
Fanti, C., et al. 2000, A\&A, 358, 499

\bibitem[Fanti et al.(2001)]{fanti01}
Fanti, C., Pozzi, F., Dallacasa, D., Fanti, R., Gregorini, L.,
Stanghellini, C., Vigotti, M. 2001, A\&A, 369, 380 

\bibitem[Fanti et al.(2004)]{cf04}
Fanti, C., et al. 2004, A\&A, 427, 465

\bibitem[Fomalont et al.(2000)]{fomalont00}
Fomalont, E.B., Frey, S., Paragi, Z., Gurvits, L.I., Scott, W.K.,
Taylor, A.R., Edwards, P.G., Hirabayashi, H. 2000, ApJS, 131, 95 

\bibitem[Giovannini et al.(2001)]{gg01}
Giovannini, G., Cotton, W.D., Feretti, L., Lara, L., Venturi, T. 2001, ApJ, 552, 508

\bibitem[Giroletti \& Polatidis(2009)]{giroletti09}
Giroletti, M., Polatidis, A.G. 2009, AN, 330, 193

\bibitem[Giroletti et al.(2010)]{giroletti10}
Giroletti, M., Reimer, A., Fuhrmann, L., Pavlidou, V. Richards, J.L. 2010, eConf Proceedings C091122 (arXiv:1001.5123)

\bibitem[Hardcastle et al.(1997)]{hardcastle97}
Hardcastle, M.J., Alexander, P., Pooley, G.G., Riley, J.M. 1997,
MNRAS, 288, 859

\bibitem[Helmboldt et al.(2007)]{helmboldt07}
Helmboldt, J.F., et al. 2007, ApJ, 658, 203

\bibitem[Jeyakumar et al.(2005)]{jeyakumar05}
Jeyakumar, S., Wiita, P.J., Saikia, D.J., Hooda, J.S. 2005, A\&A, 432, 823

\bibitem[Kameno et al.(2000)]{kameno00}
Kameno, S., Horiuchi, S., Shen, Z.-Q., Inoue, M., Kobayashi, H., Hirabayashi, H., Murata, Y. 2000, PASJ, 52, 209

\bibitem[Labiano et al.(2006)]{labiano06}
Labiano, A., Vermeulen, R.C., Barthel, P.D., O'Dea, C.P., Gallimore,
J.F., Baum, S., de Vries, W. 2006, A\&A, 447, 481

\bibitem[Labiano et al.(2007)]{labiano07}
Labiano, A., Barthel, P.D., O'Dea, C.P., de Vries, W.H., P\'erez, I.,
Baum, S.A. 2007, A\&A, 463, 97 

\bibitem[Lawrence et al.(1996)]{lawrence96}
Lawrence, C.R., Zucker, J.R., Readhead, A.C.S., Unwin, S.C., Pearson,
T.J., Xu, W. 1996, ApJS, 107, 541 

\bibitem[Lenc et al.(2008)]{lenc08}
Lenc, E., Garrett, M.A., Wucknitz, O., Anderson, J.M., Tingay,
S.J. 2008, ApJ, 673, 78

\bibitem[Morganti et al.(2004)]{morganti04}
Morganti, R., Oosterloo, T.A., Tadhunter, C.N., Vermeulen, R.,
Pihlstr\"om, Y.M., van Moorsel, G., Wills, K.A. 2004, A\&A, 424, 119 

\bibitem[Murgia et al.(1999)]{murgia99}
Murgia, M., Fanti, C., Fanti, R., Gregorini, L., Klein, U., Mack,
K.-H., Vigotti, M. 1999, A\&A, 345, 769 

\bibitem[Murgia(2003)]{murgia03}
Murgia, M. 2003, PASA, 20, 19

\bibitem[Murphy et al.(1993)]{murphy93} 
Murphy, D.W., Browne, I.W.A., Perley, R.A. 1993, MNRAS, 264, 298

\bibitem[O'Dea et al.(1991)]{odea91}
O'Dea, C.P., Baum, S.A., Stanghellini, C. 1991, ApJ, 380, 66

\bibitem[O'Dea(1998)]{odea98}
O'Dea, C.P. 1998, PASP, 110, 493

\bibitem[Orienti et al.(2004)]{mo04}
Orienti, M., Dallacasa, D., Fanti, C., Fanti, R., Tinti, S.,
Stanghellini, C. 2004, A\&A, 426, 463 

\bibitem[Orienti et al.(2007)]{mo07}
Orienti, M., Dallacasa, D., Stanghellini, C. 2007, A\&A, 461, 923

\bibitem[Orienti \& Dallacasa(2008)]{mo08}
Orienti, M., Dallacasa, D. 2008, A\&A, 487, 885

\bibitem[Orienti \& Dallacasa(2010)]{mo10}
Orienti, M., Dallacasa, D. 2010, MNRAS, 406, 529

\bibitem[Owsianik \& Conway(1998)]{owsianik98}
Owsianik, I., Conway, J.E. 1998, A\&A, 337, 69

\bibitem[Pacholczyk(1970)]{pacho70} 
Pacholczyk, A.G. 1970, Radio Astrophysics, Freeman and Co., San
Francisco 

\bibitem[Peacock et al.(1981)]{peacock81}
Peacock, J.A., Perryman, M.A.C., Longair, M.S., Gunn, J.E., Westphal,
J.A., 1981, MNRAS, 194, 601

\bibitem[Peacock \& Wall(1981)]{pw81} 
Peacock, J.A., Wall, J.V. 1981, MNRAS, 194, 331

\bibitem[Peck \& Taylor(2000)]{peck00}
Peck, A.B., Taylor, G.B. 2000, ApJ, 534, 90

\bibitem[Peck \& Taylor(2001)]{peck01}
Peck, A.B., Taylor, G.B. 2001, ApJ, 554, 147

\bibitem[Polatidis \& Conway(2003)]{polatidis03}
Polatidis, A.G., Conway, J.E. 2003, PASA, 20, 69

\bibitem[Polatidis(2009)]{polatidis09}
Polatidis, A.G. 2009, AN, 330, 149

\bibitem[Readhead(1994)]{readhead94}
Readhead, A.C.S. 1994, ApJ, 426, 51

\bibitem[Rossetti et al.(2006)]{rossetti06}
Rossetti, A., Fanti, C., Fanti, R., Dallacasa, D., Stanghellini,
C. 2006, A\&A, 449, 49 

\bibitem[Saikia et al.(2003)]{saikia03}
Saikia, D.J., Jeyakumar, S., Mantovani, F., Salter, C.J., Spencer,
R.E., Thomasson, P., Wiita, P.J. 2003, PASA, 20, 50 

\bibitem[Salter et al.(2010)]{salter10}
Salter, C.J., Saikia, D.J., Minchin, R., Ghosh, T., Chadola, Y. 2010,
ApJ, 715, 117

\bibitem[Siemiginowska et al.(2005)]{aneta05}
Siemiginowska, A., Cheung, C.C., LaMassa, S., Burke, D.J., Aldcroft,
T.L., Bechtold, J., Elvis, M., Worrall, D.M. 2005, ApJ, 632, 110
 
\bibitem[Snellen et al.(2000)]{snellen00}
Snellen, I.A.G., Schilizzi, R.T., Miley, G.K., de Bruyn, A.G., Bremer,
M.N., R\"ottgering, H.J.A. 2000, MNRAS, 319, 445

\bibitem[Stanghellini et al.(1998)]{stanghellini98}
Stanghellini, C., O'Dea, C.P., Dallacasa, D., Baum, S.A., Fanti, R.,
Fanti, C. 1998, A\&AS, 131, 303 

\bibitem[Stickel et al.(1994)]{stickel94}
Stickel, M., Meisenheimer, K., K\"uhr, H. 1994, AAS, 105, 211

\bibitem[Stickel \& K\"uhr(1996)]{stickel96}
Stickel M., K\"uhr, H. 1996, A\&AS, 115, 11

\bibitem[Taylor et al.(1996)]{taylor96}
Taylor, G.B., Readhead, A.C.S., Pearson, T.J. 1996, ApJ, 463, 95

\bibitem[Vermeulen et al.(2003)]{vermeulen03}
Vermeulen, R.C., Pihlstr\"om, Y.M., Tschager, W., et al. 2003, A\&A,
404, 861
 
\bibitem[Veron-Cetty \& Veron(2006)]{veron06}
Veron-Cetty, M.-P., Veron, P. 2006, A\&A, 455, 773

\bibitem[Wilkinson et al.(1994)]{wilkinson94}
Wilkinson, P.N., Polatidis, A.G., Readhead, A.C.S., Xu, W., Pearson,
T.J. 1994, ApJ, 432, 87

\bibitem[Willott et al.(1998)]{willott98}
Willot, C.J., Rawlings, S., Blundell, K.M., Lacy, M., 1998, MNRAS, 300, 625

\bibitem[Xu et al.(1994)]{xu94}
Xu, W., Lawrence, C.R., Readhead, A.C.S., Pearson, T.J. 1994, AJ, 108, 395

\bibitem[Xu et al.(1995)]{xu95}
Xu, W., Readhead, A.C.S., Pearson, T.J., Polatidis, A.G., Wilkinson,
P.N. 1995, ApJS, 99, 297

\bibitem[Zirbel \& Baum(1995)]{zirbel95}
Zirbel, E.L., Baum, S.A. 1995, ApJ, 448, 521

\end{thebibliography}
\end{document}